\documentclass[onecolumn,numberedappendix,apj]{emulateapj}
\usepackage[utf8]{inputenc}
\usepackage{amsmath}
\usepackage{amsfonts}
\usepackage{amssymb}
\usepackage{graphicx}
\slugcomment{{\sc Submitted to ApJ:} November 22, 2016}

\shorttitle{Dense Basis SED fitting}
\shortauthors{Iyer \& Gawiser}

\pdfoutput = 1

\begin{document}

\title{Reconstruction of Galaxy Star Formation Histories through SED Fitting:\\ The Dense Basis Approach}

\author{Kartheik G. Iyer\altaffilmark{1} and Eric Gawiser\altaffilmark{1}}

\affil{Dept. of Physics and Astronomy, Rutgers}

\begin{abstract}

We introduce the Dense Basis method for Spectral Energy Distribution (SED) fitting. It accurately recovers traditional SED parameters, including M$_*$, SFR and dust attenuation, and reveals previously inaccessible information about the number and duration of star formation episodes and the timing of stellar mass assembly, as well as uncertainties in these quantities. This is done using basis Star Formation Histories (SFHs) chosen by comparing the goodness-of-fit of mock galaxy SEDs to the goodness-of-reconstruction of their SFHs. We train and validate the method using a sample of realistic SFHs at $z =1$ drawn from stochastic realisations, semi-analytic models, and a cosmological hydrodynamical galaxy formation simulation. The method is then applied to a sample of 1100 CANDELS GOODS-S galaxies at $1<z<1.5$ to illustrate its capabilities at moderate S/N with 15 photometric bands. Of the six parametrizations of SFHs considered, we adopt linear-exponential, bessel-exponential, lognormal and gaussian SFHs and reject the traditional parametrizations of constant (Top-Hat) and exponential SFHs. We quantify the bias and scatter of each parametrization. $15\%$ of galaxies in our CANDELS sample exhibit multiple episodes of star formation, with this fraction decreasing above $M_*>10^{9.5}M_\odot$. About $40\%$ of the CANDELS galaxies have SFHs whose maximum occurs at or near the epoch of observation. The Dense Basis method is scalable and offers a general approach to a broad class of data-science problems.
\end{abstract}

\keywords{galaxies: star formation | galaxies: evolution | techniques: photometric}

\section{Introduction}
\label{sec:intro}

The integrated light of a galaxy offers a vast amount of information. When measured with sufficient precision and suitably analysed, the Spectral Energy Distribution (SED) offers insights about a galaxy's composition from its birth to its time of observation \citep{acquaviva2011sed,conroy2010propagation}. This can be used to estimate the galaxy's star formation rate as a function of time, which traces its evolution and merger history \citep{moped, vespa, beagle, prospector}. Combined with other observations, this provides valuable knowledge of cosmic structure formation.

Existing methods of SED fitting use a variety of sophisticated techniques. These include inversion methods \citep{moped}, bayesian codes for estimating uncertainties and covariances \citep{acquaviva2015simultaneous, beagle},  machine learning methods with training sets \citep{leistedt2016data}, and template-based models \citep{bolzonella2000photometric}. To search the large parameter spaces of the variables in consideration, Markov Chain Monte Carlo (MCMC) methods have become increasingly popular. 

These advances have been necessitated by the increasing detail provided by theory, and the expanding size of galaxy catalogues available through surveys.  A large amount of (spectro)photometric data of unprecedented quality will be generated in upcoming surveys, like LSST \citep{lsst}, HETDEX \citep{hetdex} and J-PAS \citep{jpas}. SDSS \citep{eisenstein2011sdss} has already measured spectrophotometry for $\sim 10^6$ objects. The HETDEX/SHELA field will cover roughly 600,000 objects with multi-band photometry and fiber spectroscopy. J-PAS will cover 9,000 square degrees with 59 filters ($ugriz$+54 narrow-band filters across optical) for $\sim 9\times 10^7$ galaxies. Large regions covered in the NIR with Euclid \citep{euclid} and WFIRST \citep{wfirst} will overlap with LSST, which leads to SEDs for $\sim 10^8$ objects by 2022, many of which will have panchromatic photometry. In keeping with the large amounts of reduced data generated by these collaborations, it is imperative that advanced methods of analysis are developed in order to gain useful information from the integrated light of the galaxies under consideration.

The star formation history (SFH) of a galaxy can sometimes be poorly constrained through different approaches to SED fitting. Typical methods assume a predetermined parametrization like constant star formation or exponentially declining star formation to estimate physical quantities of interest like the stellar mass, star formation rate (SFR), or the time at which the galaxy started forming stars. A few approaches instead seek to reconstruct the SFH from the data, using methods that include reducing the dimensionality of the parameter space using data compression methods \citep{moped}, fine-binning the interval that makes the maximum contribution to flux \citep{vespa}, mapping the discretized-time photometric fitting to a linear inversion problem \citep{dye}, or comparing against a large basis of realistic model SEDs using a Bayesian method \citep{pacifici}. In the current work, we aim to show that using a well-motivated basis allows us to reconstruct robust star formation histories from galaxy SEDs. 

The paper is organized as follows: in \S \ref{sec:2}, we introduce the  Dense Basis formalism of SED Fitting, and how it can be applied to the specific problem of reconstructing SFHs, including the motivation for a particular choice of basis and the fitting procedure with a particular basis set. We describe the training of the atlas using different sources of realistic SFHs: SAMs, Hydrodynamic simulations, and stochastic SFHs in \S \ref{sec:traingofgor}. In \S \ref{sec:valsec} we validate the method on both synthetic SEDs from the SAMs as well as real SEDs from the CANDELS GOODS-S field. We then present results in \S \ref{sec:results} including the number of episodes of star formation in the galaxy's past and constraints on the timing and duration of star formation activity, quantities that were previously inaccessible through SED fitting.  In \S \ref{sec:discussion}, we discuss biases introduced by adopting single parametrizations of SFHs, compare with other SFH reconstruction methods, and mention the application of the Dense Basis method to larger datasets.

\section{The Dense Basis Formalism}
\label{sec:2}

\begin{figure}[ht!]

\begin{center}

\includegraphics[width=500px]{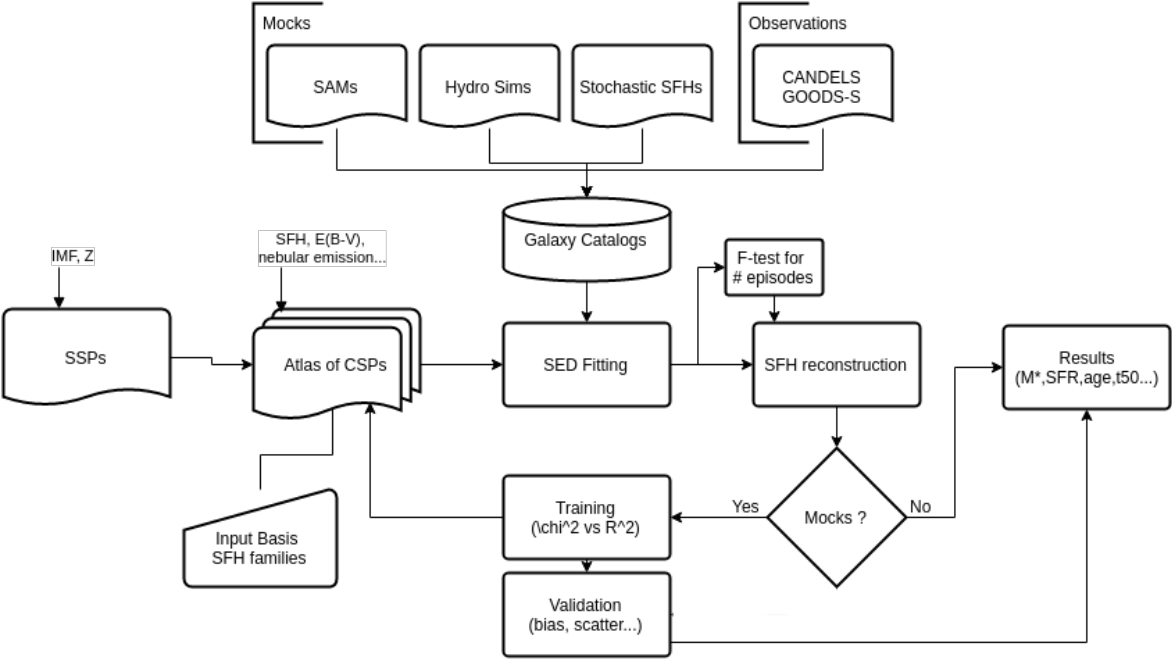}
\caption{Schematic workflow describing the current implementation of the dense basis method to the reconstruction of Star Formation Histories through photometric SED fitting. 
}
\label{fig:2}
\end{center}
\end{figure}

The Dense Basis SED fitting method reconstructs Star Formation Histories (SFHs) of individual galaxies using an atlas comprised of SEDs corresponding to well motivated families of SFHs that effectively cover the space of all physical SFHs\footnote{While an expansion using an infinite number of polynomials or a fourier decomposition  would provide a true basis in the sense of spanning the space of all possible curves, our basis functions only do so approximately; however, since they can reconstruct any star formation history to the level of precision attainable with spectrophotometric data, they provide an effective basis.}. It does so by training the atlas on mock catalogs prior to fitting the full dataset. This allows us to use the reconstructed SFHs to perform novel analyses and to tackle problems that were previously intractable with SED fitting, such as estimating the number and duration of star formation episodes in a galaxy's past. To avoid any bias due to choice of prior, the method is currently implemented in a frequentist manner. In this section, we briefly describe the Dense Basis methodology and training of the basis set. 
An overview of the process is described in Figure$~$\ref{fig:2}.

\subsection{A well motivated basis of SFHs}
\label{sec:sfhcondns}

The collection of multiple families of well motivated SFHs and their corresponding SEDs with which we fit galaxies; henceforth atlas of SEDs and SFHs, should be designed to utilise the Dense Basis method to its full potential. The choice of appropriate families of functions to best describe the formation of stars in the galaxy in SFH space ($SFR ~vs~ t$) determines how the SED-fitting procedure encodes realistic star formation. We employ seven major considerations in the choice of basis that should be satisfied for every functional family under consideration:  

\begin{itemize}

\item \textbf{Physically or empirically motivated:} The functional form of the SFH needs to be realistic, arising either from statistical analysis of star formation in model galaxies, or deduced from observed galaxies. For the latter, as in \citet{gladderslognormal}, skewed distributions such as linear rise followed by exponential decline and lognormal arise in physical processes restricted to non-negative domains. In this case, SFHs should also satisfy $SFR|_{t=0} \sim 0$ at the Big Bang.

\item \textbf{Robustness of reconstruction:} The family of basis SFHs should be chosen such that a good fit in an SED space ($[F_\nu,\lambda]$) should correspond to a good reconstruction in SFH space ($[SFH(t),t]$). This correspondence can be tested in various ways and could potentially be different for different datasets since the representative form of the SFH could differ across epochs. It is a useful metric for eliminating SFH families that fit SEDs well but yield biased SFH results, such as exponentially declining SFH parametrizations, which describe star formation reasonably well at recent times, but bias quantities such as Age and $t_{50}$, the lookback time at which the galaxy accumulates 50\% of its observed mass, \citep{pacifici}. Analogous to isochrone synthesis and matrix inversion methods, this is possible since the SEDs are piecewise linear in their dependence on the SFH and can be decomposed into multiple representations using different functional families.

\item \textbf{Dense in SFH space:} To avoid degeneracies and biases, (i.e., to better reveal the local minima of the likelihood surface in parameter space), we need to ensure the basis is sufficiently dense in the space of n-parameter curves spanning $SFR(t)$ in the interval $t\in [0,t_{obs}]$.

\item \textbf{Minimal number of parameters:} The number of parameters used to describe the functional form of the SFH basis functions will determine the amount of data compression possible in reconstructing the spectrum of the galaxy from its best-fit coordinates in parameter space: $SED(M_*, SFR(t), Z(t), A_v,...)$. For the present application, we model the star formation history as a sum of star formation basis functions, each needing three parameters to describe each reconstructed episode of star formation, the timing of the peak, the timescale, and the stellar mass formed..

\item \textbf{Temporally consistent:} The families should be chosen such that they produce consistent results for an SFH, independent of when the galaxy is observed, within uncertainties.

\item \textbf{Positive definite:} Any functional used to describe the SFH should be positive definite, since $SFR(t) \geq 0,~t\in [0,t_{obs}]$, which allows us to extract physical information from multicomponent solutions to the reconstructed SFH, as opposed to methods like PCA \citep{ferreras2006principal} or piecewise-linear matrix inversion \citep{dye}, which need regularization to yield physical solutions.
 
\item \textbf{Robust to noise:} The atlas spans the space of physically motivated SFHs, but not the space of all possible SEDs. This makes it robust to noise in the sense that distortions due to noise that are not accessible through the physically motivated families of SFHs under consideration do not bias the fits, as described in Appendix.\ref{sec:noise_robust}.

\end{itemize}

\begin{figure}[ht!]
\plotone{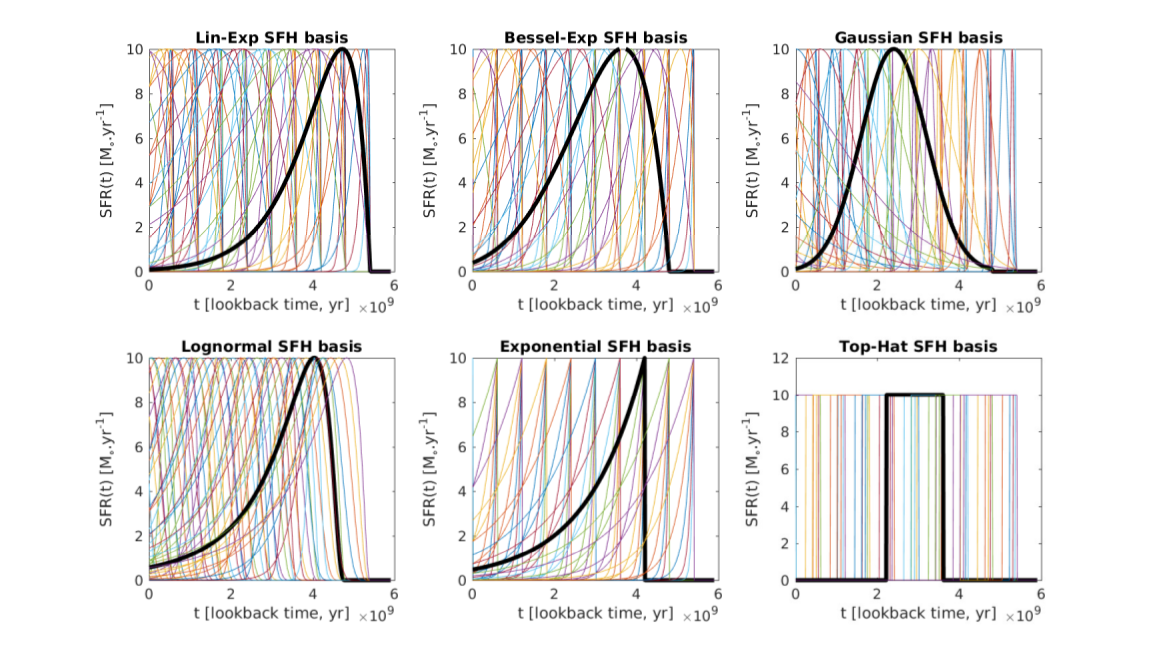}
\caption{Representative examples of the SFHs at z=1 for the different functional families described in \S\ref{sec:formulation}. The full atlas for the Dense Basis method is constructed using all physical combinations of the Linexp, Besselexp, Gaussian and lognormal families. A representative curve is shown in bold for each family.}
\label{fig:thedensebasissfhs}
\end{figure}

We describe a few of the 2-parameter families of curves for the current analysis. An overall normalization corresponding to the stellar mass acts as a third parameter. A visual representation of these families is shown in Figure \ref{fig:thedensebasissfhs}.

\begin{enumerate}

\item \textbf{Top-Hat:} Historically, simple stellar populations (SSPs) assumed that a galaxy's stellar population formed in a single instantaneous burst \citep{tinsley1980evolution}. An improvement over that was the extension to constant star formation (CSF) from a start time through the time of observation at a fixed rate. Here we use a two-parameter version of this parametrization, with a start time and a width\footnote{Since this is a positive definite version of the Top-Hat wavelet, this illustrates the possibility of extending our method to a wavelet basis.}. This is also useful for comparison with quantities in the literature computed using CSF histories, which correspond to setting $\tau \geq t_{obs}-t_0$.
\begin{equation}
SFR(t,t_0,\tau) = \Theta (t-t_0) (1-\Theta(t-t_0-\tau))
\end{equation}
where $\Theta(t)$ denotes the Heaviside function with $\Theta(t) = 1$ for $t \geq 0$ and $\Theta(t) = 0$ for $t<0$, $t_0$ is the time at which star formation starts, and $\tau$ is the width of the Top-Hat. 

\item \textbf{ESF:} Exponentially declining star formation rates, a parametrization that performs well for local ellipiticals and for comparison with older literature with $t_0$ the time at which star formation starts and $\tau$ the rate constant of the exponential decline.
\begin{equation}
SFR(t,t_0,\tau) = \Theta (t-t_0) \exp (-\frac{(t-t_0)}{\tau})
\end{equation}

\item \textbf{Linexp:} The delayed exponential \citep{lee2010estimation, gavazzi, behroozi2010comprehensive} with an additional parametrised start-time $t_0$ (henceforth $Linexp$) giving the time at which star formation starts and $\tau$ setting the width of the episode of star formation.
\begin{equation}
SFR(t,t_0,\tau) = \Theta(t-t_0) ((t-t_0)/\tau) e^{-(t-t_0)/\tau}
\end{equation}

\item \textbf{Gaussian:} A parametrization that is useful for describing symmetric episodes of star formation, where $t_{peak}$ is the time at which star formation peaks and $\tau$ is the standard deviation, which sets the width of the episode of star formation.
\begin{equation}
SFR(t,t_{peak},\tau) = \exp\left( \frac{-(t-t_{peak})^2}{2\tau^2} \right)
\end{equation}

\item \textbf{Lognormal:} \citep{gladderslognormal,dressler}. A two-parameter statistical distribution that appears in many physical processes, $t_0$ is the time at which star formation starts and $\tau$ sets the width of the episode of star formation.
\begin{equation}
SFR(t,t_0,\tau)  = \Theta(t-t_0) \frac{1}{t} \exp \frac{-(\ln (t-t_0))^2}{2\tau^2}
\end{equation}

\item \textbf{Besselexp:} Bessel-function rise, followed by exponential decline (henceforth $Besselexp$). The order of the Bessel function of the first kind, $\nu$ determines when the SFR peaks\footnote{Although there is no closed form expression for this, it can be easily determined from a lookup table for the zeros of $J'_\nu (t/\tau)$ and to linear approximation is $t_{peak} \sim 1.5(10^8) \nu$ Yr}, and $\tau$ sets the width of the episode of star formation. 
\begin{equation}
SFR(t,\nu,\tau) = J_\nu(t/\tau)e^{-t/\tau} + \alpha t
\end{equation}
We add a linear piece such that $\alpha t_{min} = - min(J_\nu (t/\tau)e^{-t/\tau})$), to ensure that the set of functions described by this family remains positive definite, while also satisfying $SFR(t=0) = 0$ at the big bang.

\end{enumerate}

\begin{figure}[ht!]
\includegraphics[width=500px]{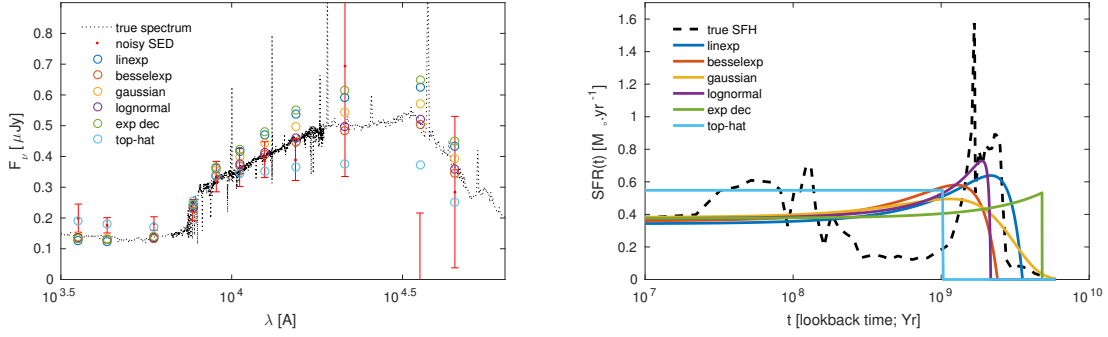}

\caption{Reconstruction of SAM mock star formation history using the six SFH parametrisations being considered as candidates for the Dense Basis method. \textit{Left panel:} Blue curve shows the true spectrum at $z = 1$. Red datapoints show the noisified SED obtained by multiplying with filter transmission curves and adding photometric noise realized from a quadrature sum of CANDELS photometric and zeropoint uncertainties (10\% for the U\_CTIO, Ks and IRAC ch1,2 bands, and 3\% for the remaining photometric bands: f435w,f606w,f775w,f850lp,f105w,f125w,f160w). Colored circles show the best fit SEDs corresponding to each reconstructed star formation history. \textit{Right panel:} Black dashed curve shows the SAM star formation history. Colored curves indicate SFR at a given lookback time at $z = 1$ for SFHs from each family that are best fits to the noisified SED. The top-hat parametrisation underestimates the stellar mass of the galaxy by $\sim 60\%$, while the exponentially declining SFH overestimates the stellar mass by $\sim 46\%$.}
\label{fig:1}

\end{figure}

These functions offer the advantages of being able to model short episodes of star formation at specific times (small t) or long periods of star formation where the rate rises and then falls (e.g., \citet{pacifici}; \citet{tomczak2016sfr}). Figure \ref{fig:1} shows a typical star formation history drawn from simulations and fits using the six families of SFHs described above. It can be seen that the standard parametrisations of constant star formation and exponentially declining star formation under and overestimate the stellar mass of the galaxy, while the other families show an improved estimation of the general trend of star formation. Additionally, the expansion of the basis to include all physically motivated combinations of single-component SFHs will allow us to describe SFHs with multiple episodes of star formation separated by periods of relative quiescence in a galaxy's SFH.

\subsection{The SED fitting problem: reconstruction of SFHs}
\label{sec:formulation}

For a Simple Stellar Population, which assumes that all of its stars form at a single lookback time (T) and with the same metallicity (Z), the luminosity at a given wavelength ($\lambda$) is simply 
\begin{equation}
L_\lambda = \int_{t_{bb} \equiv 0}^{t_{obs}} dt' L_\lambda^{SSP} (t_{obs}-t', Z)\delta(T-t_{obs}+t') = L_\lambda^{SSP} (t_{obs} - T,Z)
\end{equation}
where $t'$ is the time since the big bang, $t_{obs}-T$ is the age of the galaxy at $t_{obs}$, and $L_\lambda^{SSP} (t,Z)$ is the spectrum giving the luminosity of an SSP of metallicity Z at age t since formation. The SSP spectrum contains assumptions for the IMF, stellar tracks, and metallicity, which we hold constant in the current study. Some of the effects of relaxing this assumption are noted in \S\ref{sec:val_sams_realZ}, and are discussed further in \S\ref{sec:compareparam}.

Generalising from Simple Stellar Populations (SSPs) to Composite Stellar Populations (CSPs), we can then represent the SED for a galaxy with a given star formation history (SFH $\equiv \psi (t)$) as an integral over all of the star formation events that occurred at different times from the birth of the universe to the time of observation. Composite stellar populations are written as a sum over a non-orthogonal set of star formation histories that satisfy the constraints outlined in \S\ref{sec:sfhcondns}, such that
\begin{equation}
\psi (t) \equiv \sum_{k} \epsilon_k \psi_k (t, \{\tau,t_0\})
\end{equation}
with $\epsilon_k \geq 0$ denoting an overall normalization corresponding to the stellar mass formed by the SFH $\psi_k(t)$. Given a basis of SFHs that spans this space, we can expand this instead as a sum over the parameter space, akin to a Fourier expansion, as,
\begin{equation}
L_\lambda = \sum_k \epsilon_k L_\lambda^k  (\psi_k (t,\{\tau,t_0\},Z)
\end{equation}
where the contribution to the luminosity from an episode of star formation $\psi_k(t,\{ \tau,t_0\})$ described by a family of curves from eq.(1-6) with the parameters $\{\tau,t_0 \}$ is given by,
\begin{equation}
L_\lambda^k = \int_{t_{bb} \equiv 0}^{t_{obs}} dt' L_\lambda^{SSP}(t_{obs}-t',Z)\psi_k(t')
\end{equation}

Dust reddening and nebular emission lines are then applied to the spectrum as described in \S\ref{sec:atlasgen}, denoted by the notation $L_{\lambda,R}$. The photometry in passband $j$ from the $k^{th}$ basis SFH $\psi_k(t)$ parametrised by $\{\tau,t_0\}$, is then given by,
\begin{align}
F_{j}^k = \frac{1}{4\pi d_L^2(1+z)} \sum_k \left( \int d\lambda  T_j (\lambda) \epsilon_k L_{\lambda,R}^i (\psi_k (t;\{\tau,t_0 \},Z))  \right)
\end{align}

Using this as a mapping from the basis of SFHs to the space of all physically motivated SEDs, we can then define a $\chi^2$ surface, which denotes the metric distance in the vector space of photometry between the observed SED and its closest match in the atlas. Finding the reconstructed SFH in the basis is then reduced to an optimization problem on the likelihood surface. For example, with a surface defined using a $\chi^2$ metric, we get

\begin{align}
\mathrm{min}(\chi^2) = \mathrm{min} \left[  \sum_j \frac{\sum_{k} \left[ \left( 4\pi d_L^2(1+z)\right)^{-1}  \int d\lambda  T_j (\lambda)\epsilon_k L_{\lambda,R}^i (\psi_k (t;\{\tau,t_0 \},Z)  -F_j^{obs} \right]^2}{\sigma_j^2} \right]
\label{eqn:minchi2prob}
\end{align}
In the following sections, we train the basis set using different mock datasets for which we can quantify both the goodness-of-fit in SED space, given by $\chi^2$ as well as the goodness-of-reconstruction in SFH space, given by $\Gamma$, defined in \S\ref{sec:traingofgor}. We choose basis functions that show sufficient correspondence between the optima of these two quantities, which lets us reconstruct SFHs in the presence of model degeneracies, systematics and instrumental noise.

\subsection{Generating the Atlas}
\label{sec:atlasgen}
In order to implement the dense-basis algorithm, it is necessary to first generate an atlas of template spectral energy distributions (SEDs) and then to use it to fit the observed SEDs. This is done as follows:

1. Basis SFHs belonging to the functional families described in \S 2.1 are generated on a grid of well-chosen discrete parameter values.

2. SEDs corresponding to these star formation histories are then generated using the isochrone synthesis code BC03. \citep{bc03}, using input parameter ranges as described in Table \ref{table:SAMinit}. 

3. \textbf{Nebular emission} is added according to the prescription in \citet{orsi2014nebular} using MAPPINGS III, a one dimensional shock and photoionization code for modelling nebular line and continuum emission. \citep{allen2008mappings}. We use in this work the precomputed HII region model grid described in \citep{kewley2001theoretical}, with the incident ionization spectra computed using Staburst99 \citep{starburst99}, at $Z_{cold~gas} = 0.2Z_{\odot}$, from which we compute the ionization parameter using,
\begin{equation}
q(Z) = 2.8\times 10^7 \left( \frac{Z_{cold}}{0.012} \right)^{-1.3}
\end{equation} 
This prescription does not add effective degrees of freedom to the atlas and could be expanded to accommodate more realistic emission in future work with higher S/N SEDs. 

4.\textbf{ Calzetti Dust Attenuation} \citep{calzetti2001dust} is applied to atlas SED spectra with discrete values of $A_v$ to extend parameter space in dust for procedures where dusty SEDs are fit, using
\begin{equation}
L_{\lambda ,R} = L_\lambda 10^{-0.4k(\lambda)A_V/R_V}
\end{equation}
where
\begin{align*}
k(\lambda) &= 2.659(-2.156+\frac{1.509}{\lambda}-\frac{0.198}{\lambda^2}+\frac{0.0011}{\lambda^3}) + R_V~~~~~~~\lambda \in [0.12,0.63] \\
&= 2.659(-1.857 + \frac{1.04}{\lambda}) + R_V~~~~~~~~~~~~~~~~~~~~~~~~~~~~~~~~\lambda \in [0.63,2.2]
\end{align*}
with $R_V = 4.05$ and the coefficients adjusted for $\lambda$ in microns. Since attenuation inferred from nebular emission lines differs from that inferred from the continuum (UV spectral slope), we use $A_{v,stars} = 0.44 A_{v,gas}$ \citep{calzetti2001dust}, where $A_{v,gas}$ is applied to both UV nebular continuum and nebular emission lines.

5. After nebular emission lines are added to the spectrum, and dust attenuation is applied, the photometry for the basis SEDs $F_{j}^k$, where $j$ denotes the photometric bands, or spectroscopic bins,  at a redshift z is given by,
\begin{equation}
F_{j} (\lambda) = \frac{1}{4\pi d_L^2 (1+z)}  \int d\lambda T_j(\lambda) L_{\lambda/(1+z),R} (\psi_k(t;\{\tau,t_0\},Z)
\end{equation}
where $T_j$ is the transmission curve of passband $j$ (the spectroscopic equivalent would be the resolution element $\Delta \lambda$ and throughput at that $\lambda$), and $d_L$ is the luminosity distance (a $d_L$ of 10 parsecs is assumed when $z=0$, as in BC03). For convenience, the flux densities are obtained as the ratio of the number of photons corresponding to the fluxes ($\lambda F_\lambda$) to the number of photons produced by a $1\mu Jy$ flat spectrum in passband $j$. This yields the observations, predictions and uncertainties in identical units. The notation $L_{\lambda,R}$ indicates that nebular emission and dust reddening  have been applied to the spectrum. 

\subsection{Choosing the number of basis functions}
\label{sec:Ftest}

In practice, galaxies rarely have sufficiently smooth star-formation histories to be perfectly fit by a functional form, as inferred from our mock datasets as well as \citet{hammer2005did, kelson, weisz2011acs,sparre2015star, diemer2017log}. In addition, considering the errors in the photometry, incomplete empirical knowledge of the mapping from SFH to SED spaces, and degeneracies between the SFH and other factors like dust and metallicity, we need to assess methods of reconstruction using multiple basis SFHs to reconstruct as close to the true SFH as possible given the quality of available data. Considering a solution to the minimization problem in Eq.\ref{eqn:minchi2prob}, we can express the Best-Fit SED as  
\begin{equation}
F_{j}^{obs} = \sum_{k=1}^{N_{basis}} \epsilon_{k} F_{j}^k(\psi_k(t)) \approx \sum_{k=1}^{N_{F}} \epsilon_k F_{j}^k(\psi_k(t))
\end{equation}
where $N_F$ is the number of components determined using the F-test, given by,
\begin{equation}
\mathcal{F}(\chi^2_{N_1},\chi^2_{N_2}) = \frac{(\chi^2_{N_1} - \chi^2_{N_2})/(d_2-d_1)}{\chi^2_{N_2}/d_2}~~~~~~\mathrm{reject~ if}~~p(\mathcal{F},d_1,d_2) < 0.5
\end{equation}
This is used to determine the number of components in the SFH space that the SED should be fit with. The F-test assesses the null hypothesis that the fit with a larger number of parameters is not a statistical improvement over a fit with a smaller number of parameters\footnote{To motivate the choice of $p=0.5$ as our bounding value, it helps to think of the case with an equal number of degrees of freedom ($N_1=N_2$), where a better statistical model has $F>1$, which corresponds to $p >0.5$. For the general case of $N_1\neq N_2$, the $p > 0.5$ cutoff provides a metric where statistical improvement is sufficient to justify the extra degrees of freedom.}, where $d_2 = N_j - 3N_2, d_1 = N_j-3N_1$ are degrees of freedom corresponding to the number of components ($N_1,N_2$) being fit with, with $N_j$ denoting the number of photometric bands. 
 
We then reconstruct the SFH according to the optimal components of the likelihood surface for the chosen $N_f$.

\subsection{Estimating Uncertainties}
\label{sec:sfhuncert}

We estimate uncertainties for the reconstructed SFHs  via a fully forward modeled frequentist approach using the likelihood surface of the fit, after rescaling the best-fit $\chi^2$ to correct for artificially low $\chi^2$ obtained for very noisy galaxies and artificially high $\chi^2$ values for the brightest galaxies. 

\begin{figure}[ht!]
\plotone{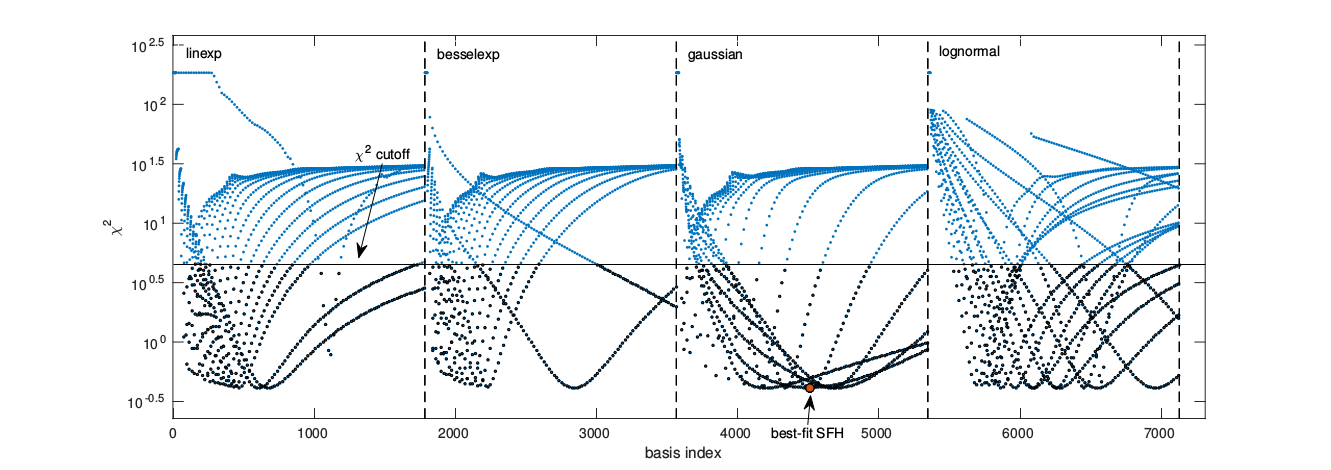}
\caption{Plot showing the full $\chi^2$ surface for an individual SAM galaxy computed using the single component basis consisting of the Linexp, Besselexp, Gaussian and Lognormal families of SFHs. Each point represents a single SFH; the SFH corresponding to the global minimum $\chi^2$ is the best-fit SFH and SFHs from all families below a threshold are used in computing the uncertainties on the reconstruction. The curves seen within each family denote $\chi^2$ for different values of $\tau$ with adjoining points differing by $\Delta t_0 \sim 0.1 dex$.}
\label{fig:chi2surf}
\end{figure}

\begin{figure}[ht!]
\plotone{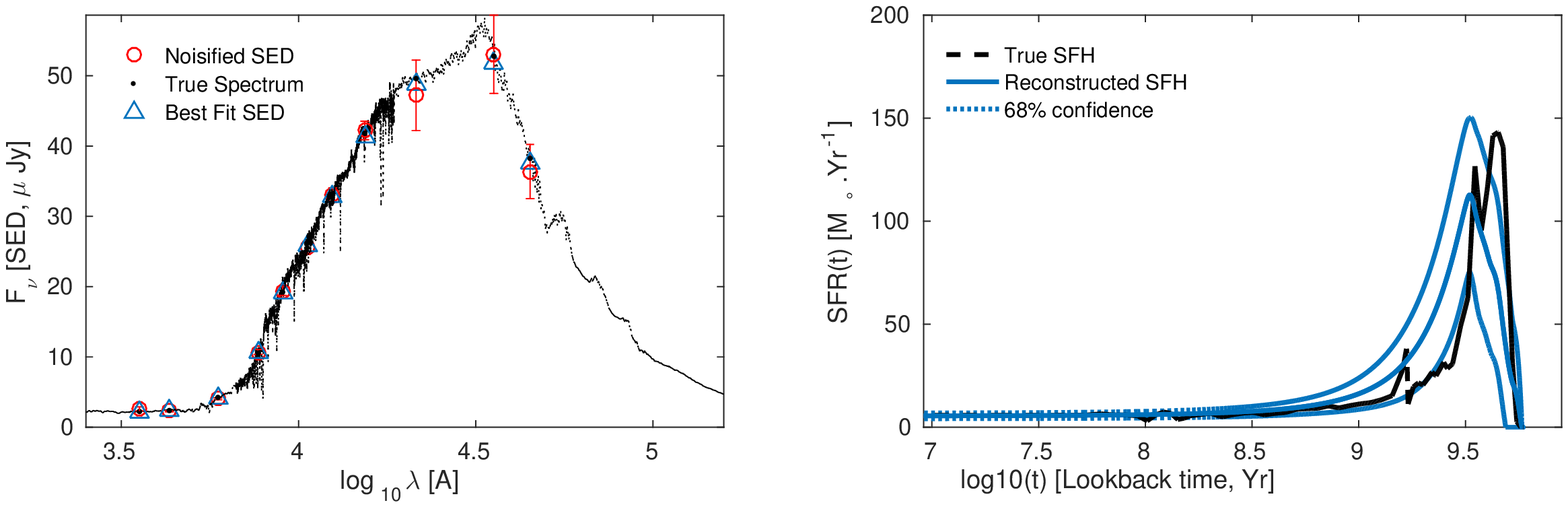}
\plotone{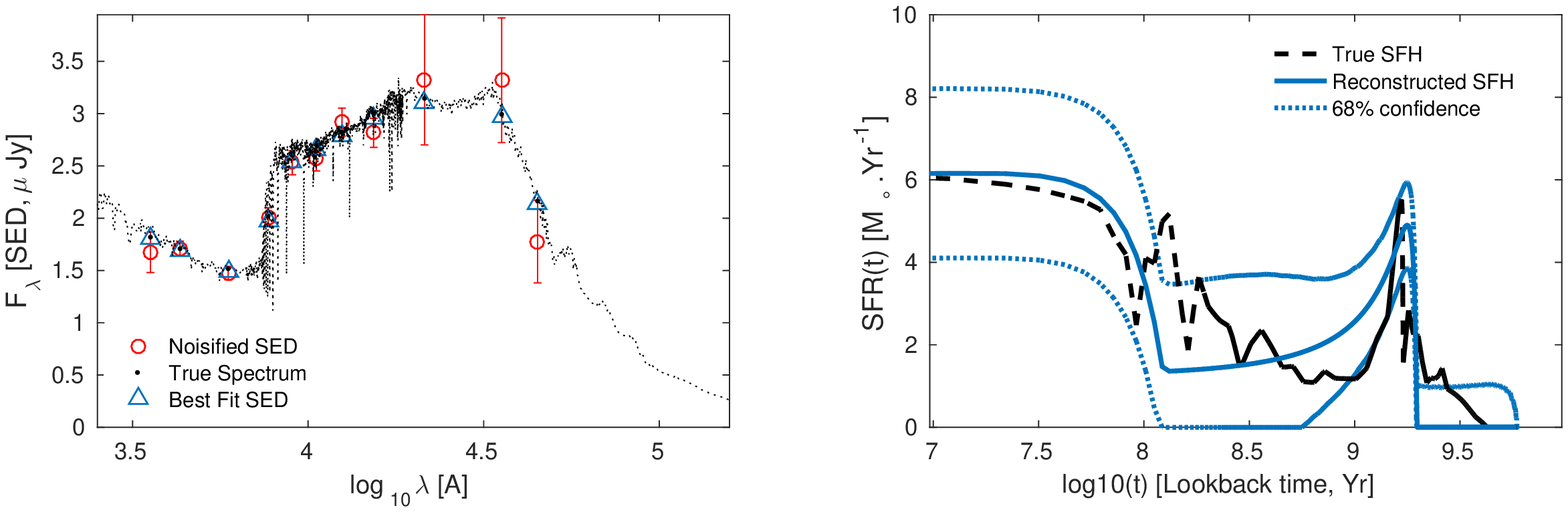}
\caption{Representative examples of SFH reconstructions with the uncertainties computed using the outlier-clipped likelihood surface, as described in \S\ref{sec:massass}. The examples show that it is possible to reasonably constrain even older episodes of star formation (top row), as well as to obtain robust uncertainties on multiple episodes (bottom row). Spectra are shown without nebular emission for clarity.}
\label{fig:uncertSFH}
\end{figure}

A subsurface of the complete likelihood surface is then obtained by imposing a cutoff using a procedure similar to \citep{avni1976energy}. We compare the SFH corresponding to each point in the subsurface to the median SFH and exclude outlier SFHs that have an excursion greater than 1.5 times the maximum value, yielding robust confidence intervals as in \citep{xie2013confidence, zhao2016robust}. The uncertainties in SFR at each point in time are then found using a distribution of the remaining acceptable SFHs. Our tests using the sample of 1200 mock SFHs show that this method robustly estimates the confidence bounds, such that for a formal $68\%$ confidence interval, the true SFH lies within the confidence interval $\sim 79\%$ of the time. We show the $\chi^2$ surface computed using this procedure for a single family in Figure \ref{fig:chi2surf} showing the best-fit SFH and threshold for uncertainties. In Figure \ref{fig:uncertSFH} we show some representative examples of the uncertainties with the top panel demonstrating the method's ability to constrain an older episode of star formation and the bottom panel showing the case of uncertainties with multiple episodes of star formation. 

\subsection{Choice of parameter space and photometric bands}
\label{sec:bandchoice}

For the initial implementation of the method in this work, we have used the Bruzual and Charlot (2003) library of stellar tracks, with the parameter space as described by Table \ref{table:SAMinit}. The Dense Basis formalism can be applied equivalently with any set of free parameters, including the set of SSP models, to use the training and validation steps to help constrain the variable parameters, as discussed in \S\ref{sec:compareparam}. All three datasets are standardized to contain a sample of 400 galaxies with the same realistic distribution of stellar mass. 

\begin{table}[ht!]
\caption{Parameter space for vetting using mock SEDs.}
\label{table:SAMinit}
\begin{center}
\begin{tabular}{ c|c c }
\hline \hline
& Parameter choice & range \\
\hline
IMF & Chabrier/Salpeter & - \\
SED generation: & BC03 & -\\
Bands fit & 11 & - \\
Tracks & Padova'94 & -\\
Metallicity & $0.2Z_\odot$ & - \\
Dust law & Calzetti & $A_v \in [0.0,2.5],R_v = 4.05$ \\
SFH form & Linexp & $\tau \in [0.014,138]$Gyr, $t_0 \in [0.02,5.9]$Gyr \\
" & Gaussian & $\tau \in [0.014,4.36]$Gyr, $t_{peak} \in [0.02,5.9]$Gyr \\
" & Besselexp & $\tau \in [1.38,4.36]$Gyr, $t_{peak} \in [0.05,5.66]$Gyr \\
" & lognormal & $\tau \in [1.38,4.36]$Gyr, $t_0 \in [0.05,5.66]$Gyr \\
" & exponential & $\tau \in [0.014,138]$Gyr, $t_0 \in [0.02,5.9]$Gyr \\
" & tophat & $\tau \in [0.014,13.8]$Gyr, $t_0 \in [0.02,5.9]$Gyr \\
\hline
\end{tabular}
\end{center}
\end{table}

For training and validation, we consider fitting 11 of the 17 CANDELS GOODS-S \citep{guo2013candels} bands: [u\_ctio, HST/ACS F435w, F606w, F775w, F850lp, HST/WFC3 F105w, F125w, F160w, VLT/HAWK-I Ks, and Spitzer/IRAC 3.6,4.5$\mu$m], excluding $u\_vimos$, F814w, F098w, and Isaac Ks for the maximum photometric orthogonality, excluding IRAC 5.8 and 8.0$\mu$m since the BC03 tracks do not account for the PAH emission that appear in those bands at z=1. Once the method was tested, we expanded to include the $u\_vimos$, F814w, F098w, and Isaac Ks bands as well, leading to fits using 15-band photometry in \S\ref{sec:val_candels}. The training is performed on the mock datasets described in \S.\ref{sec:traingofgor}, the validation is performed using both the mocks datasets as well as the CANDELS sample for which SpeedyMC results are available. Finally, the results are compiled using the full CANDELS sample at $1<z<1.5$.

We present results at z=1 in the current work since it allows us to analyse rest-UV information that comes into the UBV bands as well as the Balmer 4000A break, while avoiding dust re-emission in the mid-IR. This choice of redshift and filter set is compatible with the BC03 SPS models while providing a moderate S/N regime in which to test the reconstruction of SFHs. The procedure can be generalized to all redshifts and is discussed in \S\ref{sec:future}. 
\vspace{36pt}

\section{Training the SFH families}
\label{sec:traingofgor}

To inform the choice of a functional form for the SFH basis, we train and validate the method with three mock datasets 

To inform the choice of a functional form for the SFH basis, we train and validate the method with three mock datasets of 400 galaxies each, drawn from Semi-Analytic models, Hydrodynamical simulations, and stochastic realisations of star formation histories. We work with multiple datasets to minimise the effect of any single training set on our choice of SFH families. Using these three mock catalogs, we look at various families of 2-parameter curves, and their combinations, to find the families that perform best at reconstructing SFHs. The atlas generated using that basis is then used to fit the real catalog. Before we go into the details of the training procedure, we first briefly describe the three datasets being used.

\subsection{Training with SAMs}
\label{sec:samtrain}

The first dataset is drawn from mock catalogs with known realistic star formation histories from state-of-the-art Semi-Analytic Models \citep{somerville2015star}.  These simulations use dark matter halo `merger trees' extracted from dissipationless N-body simulations in a $\Lambda$CDM universe \citep{klypin2011dark} to determine the masses of dark matter halos collapsing at a given epoch, following which halos merge to form larger structures. In this framework, SAMs use analytic recipes to model the radiative cooling of gas, suppression of gas infall, and cooling due to the presence of a photoionizing background, collapse of cold gas to form a rotationally-supported disk, conversion of cold gas into stars, and feedback and chemical enrichment from massive stars and supernovae. 
A more recent generation of SAMs also includes prescriptions for the growth of supermassive black holes and the impact of the energy they release on galaxies and their surroundings \citep{croton2006many, bower2006breaking, somerville2008semi}. Recent comparisons have shown that SAMs produce similar predictions for fundamental galaxy properties to those of numerical hydrodynamic simulations, perhaps because of the common framework of $\Lambda$CDM, which dictates gravitationally driven gas accretion rates \citep{somerville2015star}. 
However, SAMs require orders of magnitude less computing time for a given volume than hydrodynamical cosmological simulations. The resulting galaxy star formation and enrichment histories are outputs. We then use these SFHs to produce SEDs to train and validate against, using a realistic mass distribution of galaxies with $M_* \geq 10^9 M_\odot$.

\subsection{Training with hydrodynamic simuations}

We also train the method against a set of SFHs obtained from the MUFASA meshless hydrodynamic simulations \citep{mufasa}, which satisfies multiple observational constraints like the stellar mass - halo mass relation \citep{behroozi, munshi2013reproducing}, the mass metallicity relation \citep{steidel2014strong, sanders2015mosdef}, and the SFR-M$_*$ relation \citep{speagle2014highly, kurczynski2016evolution}. The star formation histories arereported as instantaneous star formation events that take place, ranging from $\mathcal{O}(10)$ to $\mathcal{O}(10^5)$ events for different galaxies. We restrict the fits to galaxies with $M_* \geq 10^9 M_\odot$, which have well defined SFHs in the simulation. To ensure the SFHs are not artificially stochastic, we generate the SFHs by convolving the instantaneous star formation events using an Epanechnikov kernel with a width of 100Myr (R.Dave, private comm.) The galaxies in MUFASA follow a realistic mass distribution that we use to sample all three mock datasets. We restrict the fits to galaxies with $M_* \geq 10^9 M_\odot$, which have well defined SFHs in the simulation. .

\subsection{Training with stochastic SFHs}
\label{sec:stoctrain}

Following the prescription of \citet{kelson}, we generate stochastic SFHs with different values for the Hurst parameter H, which quantified the autocorrelation of the SFH. A value of $H=0.5$ corresponds to a random walk in $\Delta SFR / \Delta t$, $H>0.5$ is correlated and $H<0.5$ is anti-correlated \citep{mandelbrot1968fractional}. This provides a sample for testing the other families against to determine possible biases. 
We generate different SFHs by varying the  Hurst exponent, which encodes the long-time correlation of the stochastic SFHs in $H\in [0.5,1]$, sampling galaxies with the same mass distribution as the SAM and hydrodynamic SFH samples. We exclude from the sample SFHs with a Hurst parameter $H<0.5$ since these do not correspond to realistic looking SFHs.

\subsection{Training procedure and results:}
\label{sec:rsquared_boxplots}

We quantify the correspondence between the goodness-of-fit in SED space and the goodness-of-reconstruction in SFH space as a metric to judge the success of each family of basis functions.

For the SED goodness-of-fit, we use $\chi^2_{SED}$, given by,
\begin{equation}
\chi^2_{SED} = \sum_j \frac{(\sum_{k=1}^{N_F} \epsilon_k F_{j}^k -F_j^{obs})^2}{\sigma_j^2}
\end{equation}
where the index k sums over the entire basis of SFHs, with a number of components determined using the F-test as described in \S\ref{sec:Ftest} and the index j sums over the photometric bands. The $\epsilon_k$ is optimized for each basis function, effectively making stellar mass the normalization. The global minima of the $\chi^2$ surface corresponds to the maximum likelihood, given by $\mathcal{L} \propto e^{-\chi^2_{SED}/2}$.

To quantitatively compare how well the families perform at reconstructing the SFHs of the galaxies in the three mock catalogs, we quantify the accuracy of reconstruction of the SFH by computing the $R^2$ statistic, given by,
\begin{equation}
R^2(\psi_{true},\psi_{rec}) = 1 - \frac{\sum_t (\psi_{true}(t) - \psi_{rec}(t))^2}{\sum_t (\psi_{true} (t) - \langle \psi_{true} (t) \rangle)^2}
\end{equation}
where $R^2$ \citep{rsquared} quantifies the amount of variance explained by the fit. \textbf{We set an ambitious goal for the reconstruction by asking if it does as well as direct fits in SFH space to the true SFHs using polynomials of the same order.} Since the  true SFHs exhibit a large amount of stochasticity, the question of good $R^2$ due to overfitting does not usually occur and is handled by the F-test in \S\ref{sec:val_sams}. We define the $R^2$ statistic in logarithmic time; since the SED is sensitive to changes in the SFR over roughly equal logarithmic intervals of time, this provides a more sensitive estimator. To handle all three datasets on the same footing, since they contain SFHs with differing amounts of structure and stochasticity, we apply a small nonparametric smoothing \citep{loess} to the SFHs. This statistic has proved to be the most robust for the current application, matching the qualitative results with other statistics, as detailed in Appendix \ref{sec:r2etc}. $R^2$ ranges from $[0,1]$, with the most successful reconstruction given by $R^2\to 1$. 
 
In the noiseless regime, most galaxies show the expected correspondence between the goodness-of-fit and goodness-of-reconstruction, especially in the regime of high likelihood ($\chi^2_{SED}/DoF < 1$).
Since we can access only $\chi^2_{SED}$ observationally, this correspondence is important since it allows us to obtain a good reconstruction for a galaxy whose SED is well fit. In order for an SFH family to be robust, we require that the SFHs for the ensemble of galaxies should be reconstructed as well as possible, comparing with direct fits to the true SFH using a polynomial with the same number of degrees of freedom.
 
\begin{figure}[ht!]
\begin{center}
\plotone{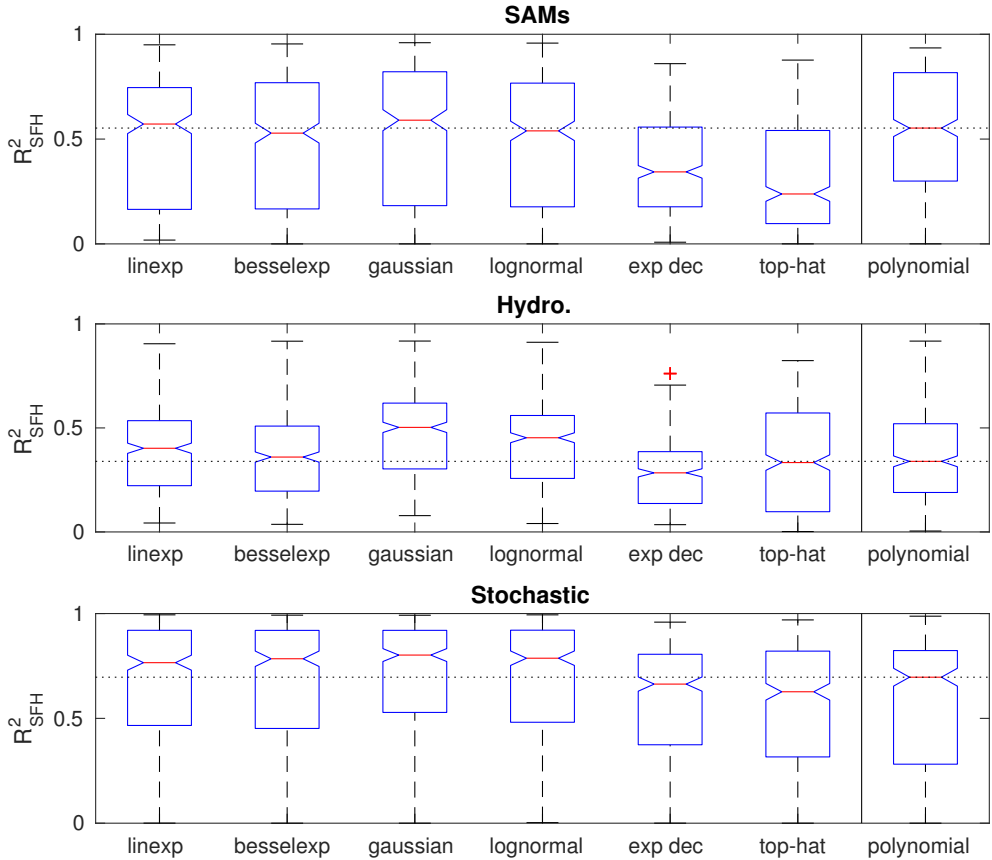}
\end{center}
\caption{Boxplots showing the accuracy of reconstruction of each SFH family to each mock dataset using the $R^2$ statistic, with the red line denoting the median and the box denoting the interquartile range. For reference, we also show the $R^2$ from direct polynomial fits to the SFH, with its median forming the horizontal dotted line. We see that on average, fits to the SEDs using the Linexp, Besselexp, Gaussian and Lognormal families perform as well as or better than direct fits to the SFHs with 3rd order polynomials.}
\label{fig:rsq_boxplot}
\end{figure}

 In Figure \ref{fig:rsq_boxplot}, we show the $R^2$ computed for each mock dataset using all six SFH families, showing that the Linexp, Besselexp, Gaussian and Lognormal families perform better overall at SFH reconstruction in comparison to the traditional parametrisations of constant and exponentially declining star formation histories. On the basis of this, we we prune our basis SFH set to retain only the Linexp, Besselexp, Gaussian and lognormal families, hereafter denoted `Best4 basis', to be used in further work.

As an additional step of validating our training statistic we examine the correspondence between $min(\chi^2_{SED})$ and a related statistic, $min(\chi^2_{SFH})$, computed using the uncertainties obtained through the method outlined in \S.\ref{sec:sfhuncert} in Appendix \ref{sec:chi2surfindgals}. We also study the possible biases that could arise in the reconstruction with a particular family, and how our choice of basis mitigates them.
\vspace{12pt}
 
\section{Validation}
\label{sec:valsec}

\subsection{Validation using three datasets: Hydrodynamic Simulations, Semi-Analytic Models and Stochastic SFHs}
\label{sec:val_sams}

Having trained our method to arrive at an optimal basis for the dataset in consideration, we now apply the SED Fitting method to the full sample of 1200 SFHs drawn from the hydrodynamic simulation, Semi-Analytic Model and the stochastic realizations.

Realistic simulated photometric noise has been applied to the mock SEDs to simulate observing conditions. The noise consists of a multiplicative factor corresponding to the zeropoint uncertainty in each band: 3\% for the space-based bands, (HST/WFC3 and HST/ACS) and 10\% for the ground-based bands (U ctio, U vimos, Isaac Ks, HawkI Ks) and IRAC Ch.1-4, as well as a photometric additive factor corresponding to the median errors in each band computed from the CANDELS dataset. With these added in quadrature to yield the $\sigma_i$ for each band $i$, simulated fluxes were drawn from a gaussian distribution $\mathcal{N}(F_\nu^i, \sigma_i^2)$.

\begin{figure}[ht!]
\begin{center}
\plotone{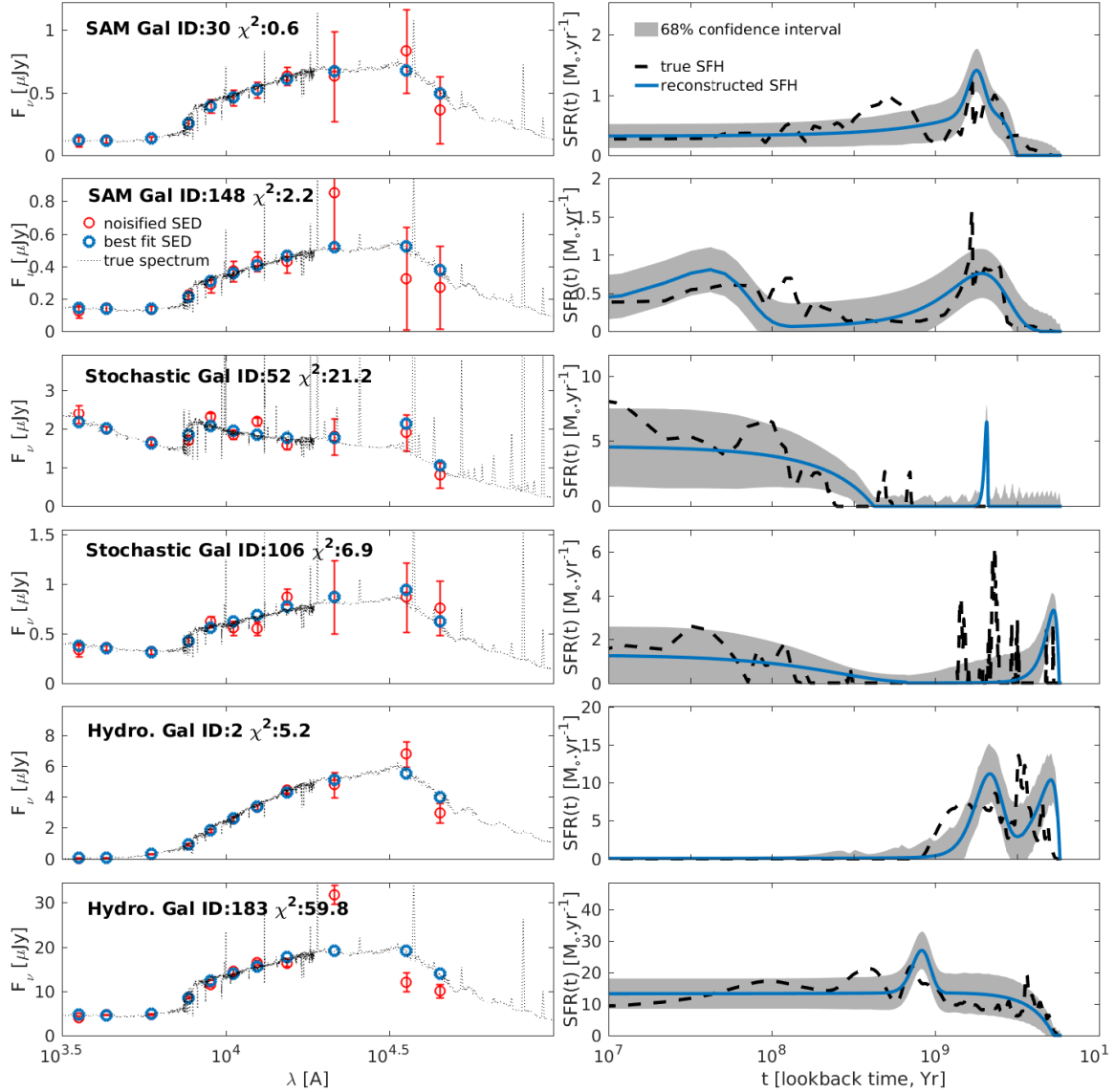}
\end{center}
\caption{The plots show a randomly drawn sample from the semi-analytic models \textbf{(Top Rows)}, Stochastic realizations \textbf{(Middle Rows)}, and hydrodynamic Simulations \textbf{(Bottom Rows)} used for the training and validation of the Dense Basis method, showing individual examples from the ensemble results shown in Figure \ref{fig:threevals}. (\textbf{Left:}) Plots show the true spectrum (black line) from the mock catalogs, their corresponding noisified photometry (red errorbars) and the best fit SED (blue open circles) using the Dense Basis method. (\textbf{Right:}) Plots show the true SFH (black dashed line) and its reconstruction (blue solid line) with 68\% confidence intervals (grey shaded region) computed using the method described in \S\ref{sec:sfhuncert}. SAM galaxy 30, is identified as a 1 episode galaxy in the current realisation of noise, constituting a false negative result of the F-test. However, many noisy realizations allow us to currently identify the second episode at $t\sim 6e8$ Yr. The episodes of star formation for the Hydro. galaxy 2 is distorted either due to noise or dust. The additional peaks in Stochastic galaxy 106 and Hydro galaxy 183 require a basis with more components. }
\label{fig:mocks_examples}
\end{figure}

We show fits using the Best4 basis that is a combination of Linexp, Besselexp, Gaussian and lognormal families, as was determined through the training step in \S\ref{sec:samtrain}. The galaxy SEDs have been fit with an atlas consisting of two component basis SFHs. This basis is constructed using all physical combinations of elements from the single episode basis and is seen to have a smaller scatter around the true values, as described in \S.\ref{sec:biasva}. In Figure \ref{fig:mocks_examples}, we show the reconstructed SFHs for two randomly selected galaxies from each mock dataset, illustrating the recovery of both recent episodes of star formation, as well as the overall trend of star formation including periods of relative quiescence.

\begin{figure}[ht!]
\begin{center}
\plotone{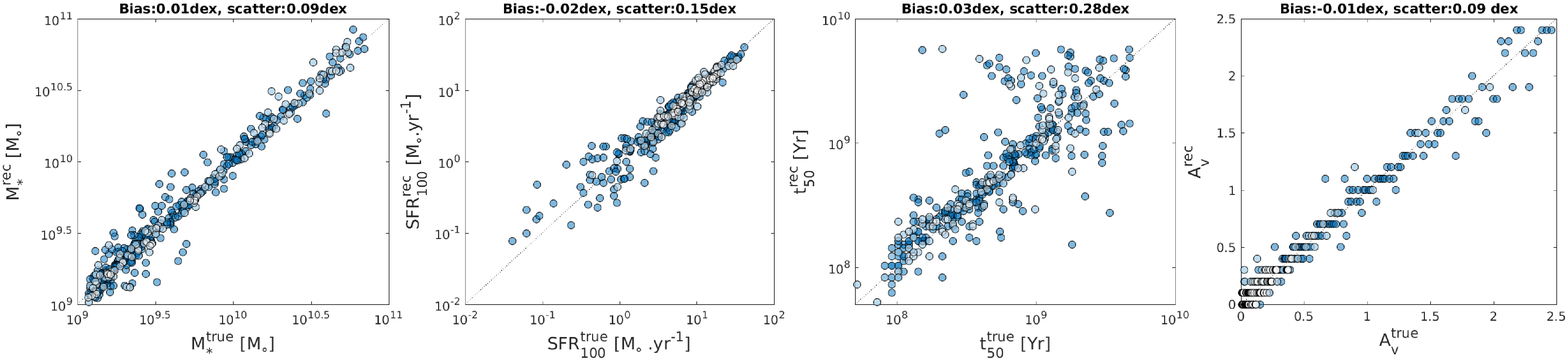}
\plotone{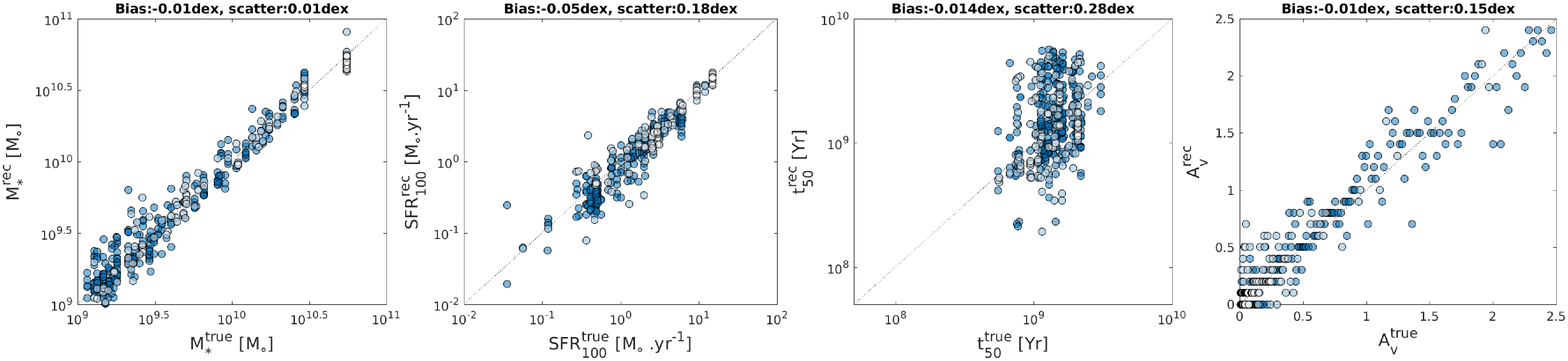}
\plotone{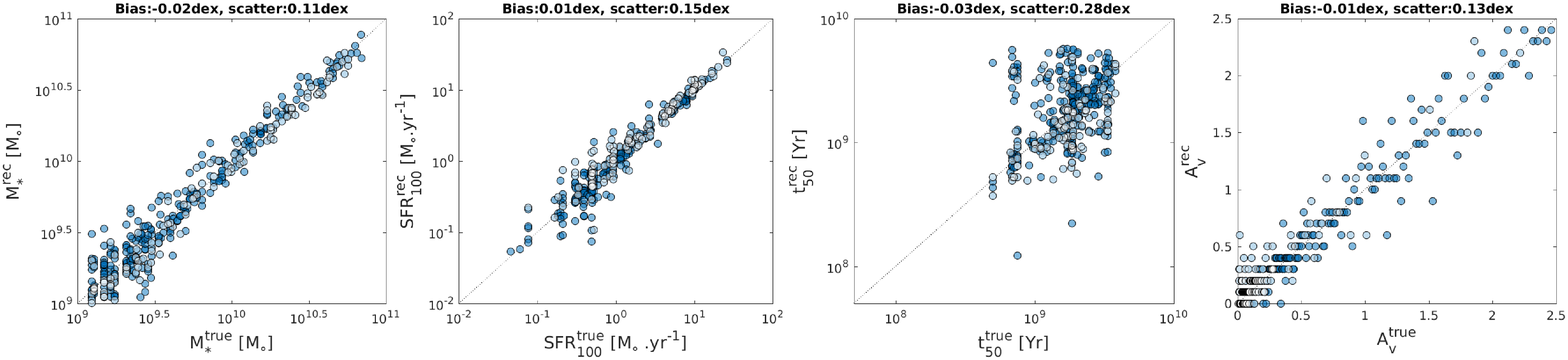}
\end{center}

\caption{We show the comparison of reconstructed against true values of the stellar mass ($M_*$), SFR$_{100}$ and $t_{50}$ for three datasets: the stochastic realizations (top row), MUFASA hydrodynamic simulations (middle row), and Semi-Analytic Model SFHs (bottom row), using the 2 episode Best4 basis fit using the Dense Basis method. For each dataset, we use a sample of 400 galaxies drawn from a realistic mass distribution. The shading indicates the likelihood of the fit, with darker shades denoting better fits to the SED. }
\label{fig:threevals}
\end{figure}

In Figure \ref{fig:threevals}, we illustrate the recovery of four physical quantities: the stellar mass ($M_*$), the star formation rate averaged over the last 100Myr ($SFR_{100}$), the lookback time over which the galaxy accreted $50\%$ of its observed stellar mass($t_{50}$), and dust extinction ($A_v$). The top row depicts the results for the stochastic realizations, the middle row for the hydrodynamic simulations, and the bottom row for the Semi-Analytic Models. All the fits show the bias to now be significantly smaller than the scatter that may occur from using a single family of SFHs, as discussed in \S\ref{sec:biasva}. The scatter in stellar mass increases at lower mass, corresponding to more noisy SEDs, while the increased scatter in SFR as compared to the values in \S.\ref{sec:biasva} is more due to the presence of dust than noise since fits without dust show a much smaller scatter of $\sim 0.05$ dex.  The 0.3 dex scatter in the reconstruction of $t_{50}$ is reasonable. However, the distributions for the SAMs and the hydrodynamic simulations look poor due to the narrow range of true values for these models with the top row being more representative of the method's performance with a broader distribution of $t_{50}$. Reconstruction of simulated dust drawn from an exponential distribution is done using an atlas containing 25 values of dust ranging from $A_v =0$ to $A_v=2.5$ using the Calzetti dust law, with reasonable scatter of $0.09-0.13 dex$ and negligible bias of $\sim 0.01 dex$.

We find that our choice of basis yields comparable good results to all three mock datasets, with the bias in the estimation of these physical quantities derived from the SFH not exceeding 0.05dex, as seen in Figure \ref{fig:threevals}. This is an important criterion to be met before the method is applied to observational data, since it lets us relax the assumption that the SFHs corresponding to the training SEDs match the actual star formation histories of galaxies at a given epoch, in favor of the slightly weaker assumption that the SFHs are drawn from a similar distribution. The ensemble results show that the reconstruction is nearly unbiased for the physical parameters of interest and can be used to extract a variety of derived quantities from the SED of distant galaxies in a robust manner.

\begin{figure}[ht!]
\begin{center}
\plotone{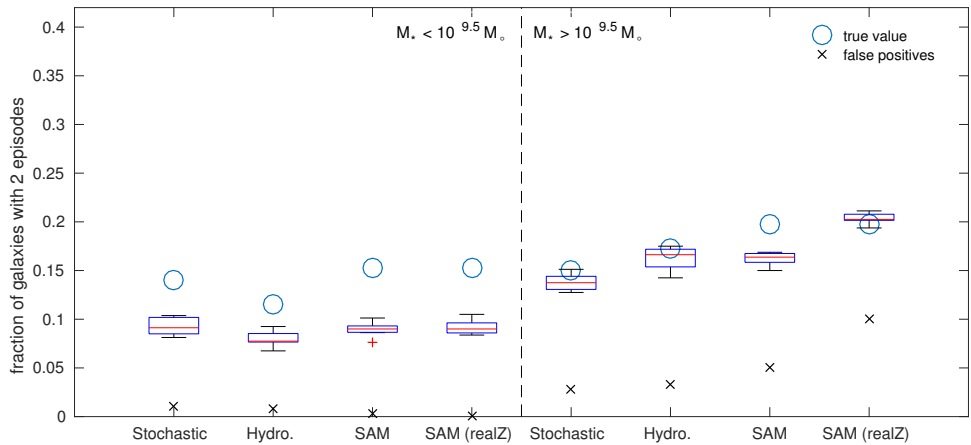}

\end{center}
\caption{We find the number of galaxies that show a statistical improvement upon being fit with a second component of star formation using the Best4 basis from \S\ref{sec:samtrain}, determine which of those correspond to a second episode of star formation, and present results for noisy realizations of four datasets of galaxies from the hydrodynamic simulations, stochastic realizations, Semi-Analytic Model, and the realistic metallicity generalization of the Semi-Analytic Model described in \S\ref{sec:val_sams_realZ}. The blue values are obtained directly from the true SFHs using a smoothing algorithm to account for stochasticity and then running a peak finder algorithm, while the black crosses quantify the number of false positive predictions using the F-test. The results are divided into two mass bins, showing that the method is reliable in predicting the number of episodes at the high mass end, due to sufficient S/N.}
\label{fig:ftest_mocks}
\end{figure}

We also present results in Figure \ref{fig:ftest_mocks} for the fraction of a sample of galaxies that are reconstructed with a second episode of star formation. For our three datasets, we perform the F-test using Eq.18, with $N_1 = 1,N_2 = 2$, and obtain the fraction of galaxies that are significantly better fit with a second component of star formation. In some cases, the second component has similar peak time and serves only to modify the SFH shape, eg. Gaussian + lognormal, with a single peak; we term these single episode SFHs. We then find the fraction of the mock galaxies that have a distinct second peak to their reconstructed SFH, which can only happen when the reconstruction prefers a second component. We compare this number to the number of galaxies in the sample whose true SFH has two episodes of star formation, computed using a peak finding routine. Since the true SFHs show a large amount of stochasticity, only the most prominent peaks with a separation greater than 100Myr are selected by smoothing over the local variations as in \S\ref{sec:samtrain} and finding the lookback times at which the SFH peaks, using $SFR'(t) = 0$ and $SFR''(t)<0$. The results are summarized as boxplots in  Figure \ref{fig:ftest_mocks}, for two mass bins chosen such that roughly half the sample lies in each mass bin. For the high mass bin, the higher S/N leads to more accurate predictions of the fraction of SEDs with more than one episode of star formation. Since our atlas is restricted to SEDs corresponding to physically motivated SFHs, we generally do not overfit the noise, as is seen by the small number of false positive results\footnote{excluding the SAM (realZ) results, which we discuss further in \S\ref{sec:val_sams_realZ}}. However, a large amount of noise makes it more difficult for the F-test to detect a statistically significant improvement to the fit. This leads to a systematic underestimation of the fraction of galaxies with more than a single episode of star formation, as shown in our results for the lower-mass galaxies. The decreased S/N also results in the fraction of fits with false negatives\footnote{which can be expressed as $f_{fn} \sim f_{true}-f_{rec}-f_{fp}$, where $f_{fn}, f_{fp}$ are the fractions of false negatives and positives, and $f_{true}, f_{rec}$ are the true and reconstructed fractions of galaxies with multiple episodes of star formation.} being higher in this mass bin.

\subsection{Validation against SpeedyMC results for CANDELS SEDs}
\label{sec:val_candels}

We now apply the method to a sample of 1100 CANDELS galaxies in the GOODS-S field at $1<z<1.5$ from \citet{kurczynski2016evolution}, for which we have physical quantities derived using SpeedyMC\footnote{a much faster version of the GalMC Markov Chain Monte Carlo algorithm \citep{acquaviva2011sed}: \\ http://www.ctp.citytech.cuny.edu/$\sim$vacquaviva/web/GalMC.html} \citep{acquaviva2011sed, acquaviva2015simultaneous} for 742 galaxies. This sample provides a good representative redshift to test the method for the recovery of SFHs at moderate S/N, as discussed towards the end of \S\ref{sec:bandchoice}. We perform the fitting with discrete values for $z$ and $A_v$, which adds some scatter to the results. For redshift, we choose bin edges at $[1.0,1.1,1.2,1.3,1.4,1.5]$, and we let $A_v$ vary from 0 to 2.5 in increments of 0.1. The purpose of this comparison is to ensure that our SED fitting code developed to implement the dense basis method, which was also used to generate mock SEDs,  does not contain circular errors. Additionally, it is a useful test to match the physical quantities that can be recovered through traditional SED fitting before presenting previously inaccessible quantities. 

\begin{figure}[ht!]
\plotone{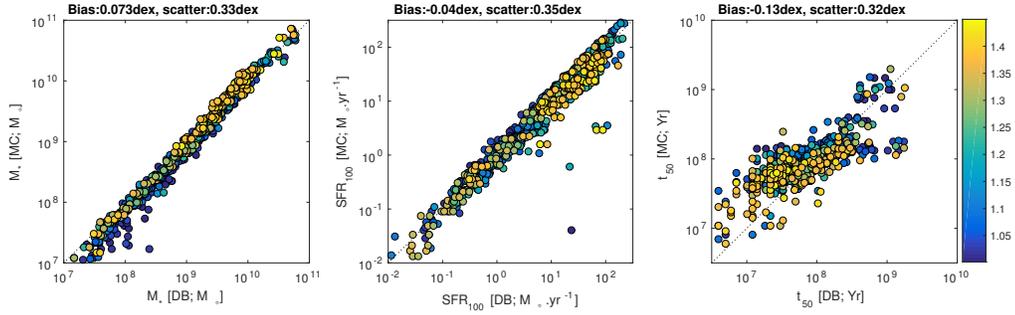}
\caption{Comparison of physical quantities derived through the Dense Basis fits against the results of the SpeedyMC Markov Chain Monte Carlo code applied to the CANDELS dataset at $1<z<1.5$ from \citet{kurczynski2016evolution}, comparing the stellar mass ($M_*$), star formation rates, and the lookback time at which the galaxies accumulated 50\% of their observed mass ($t_{50}$). The color table indicates the redshifts of the galaxies being fit.}
\label{fig:candelsfits}
\end{figure}

In order to make the comparison as consistent as possible, we match the initial conditions of the fitting procedure to the SpeedyMC parameter space, as summarized in Table \ref{table:DBMC}, and limit our SFH basis to single-component Linexp curves. In Figure \ref{fig:candelsfits}, we show the results comparing our fits to the SpeedyMC results for the stellar mass ($M_*$), SFR$_{100}$ and $t_{50}$. The slight bias in $t_{50}$ could be due to the difference in the way the two codes implement nebular emission. The colorbars denote the spectroscopic redshifts corresponding to the observed galaxies.

\begin{table}[ht!]
\caption{Comparison of Dense Basis and SpeedyMC parameter spaces used in \S\ref{sec:val_candels}}
\label{table:DBMC}
\begin{center}
\begin{tabular}{ c|c c }
\hline \hline
& Dense basis & SpeedyMC \\
\hline
IMF & Salpeter & Salpeter \\
GoF: & $\chi^2$ & $\chi^2$ \\
SPS: & BC03 & BC03 \\
Bands fit & $\leq 17$ & $\leq 17$ \\
Metallicity & $0.2Z_\odot$ & $0.2Z_\odot$ \\
Dust law & Calzetti & Calzetti \\
Nebular emission & MAPPINGS III & custom \\
SFH form & Linexp & Linexp \\
\hline
\end{tabular}
\end{center}
\end{table}

\subsection{Validation against SAM SEDs with multiple metallicities:}
\label{sec:val_sams_realZ}
We address a final possible source of systematic bias in the fits: the assumption of a single metallicity ($0.2Z_\odot$) in building the atlas and performing the fits at $z \simeq 1$. To take into account the distribution of metallicities found in real galaxies, we go back to the SAM SFHs and consider the individual metallicity components of the overall SFHs. We generate spectra corresponding to each of these metallicities using six values of metallicity available for the Padova'94 tracks in BC03, given by $Z = [0.0001,0.0004,0.004,0.008,0.02,0.05]$. Using this procedure, we obtain spectra corresponding to the SFH in each metallicity bin and use a weighted sum to obtain SEDs corresponding to galaxies with realistic metallicity histories, which we denote as SAM (realZ). We then fit these SEDs with our single-metallicity basis to test how robust our fits are at $z=1$.

\begin{figure}[ht!]
\begin{center}
\includegraphics[width=108px]{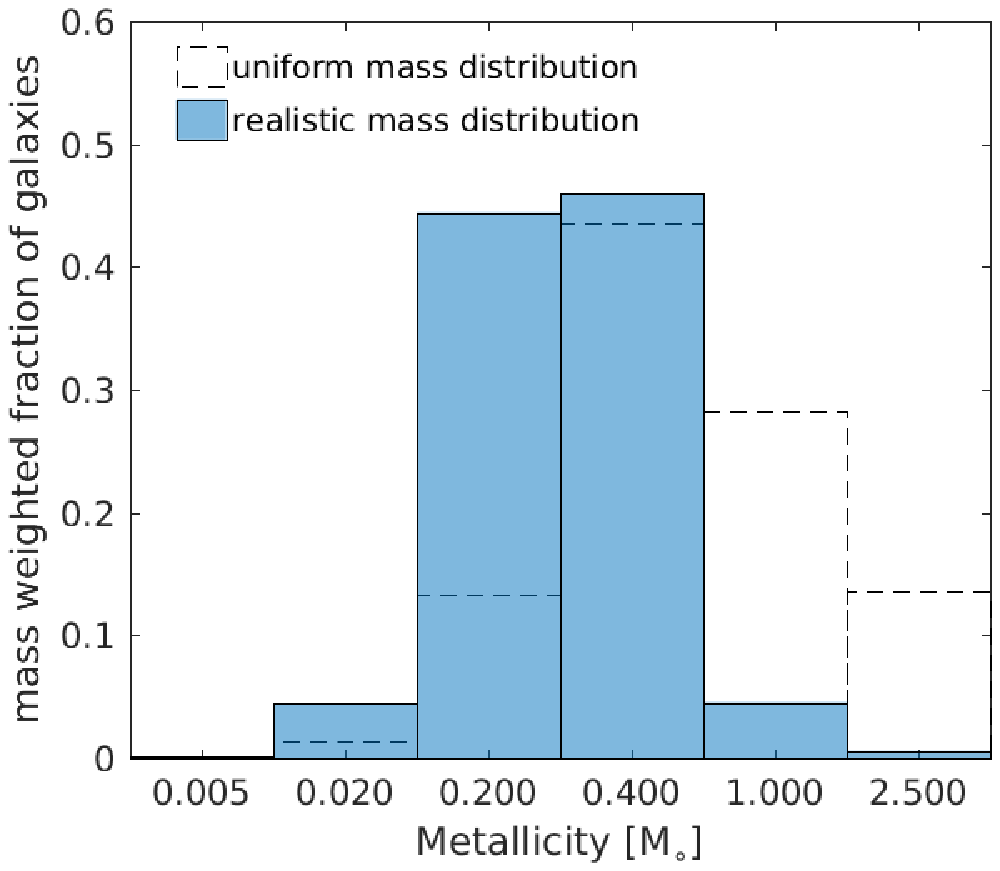}
\includegraphics[width=375px]{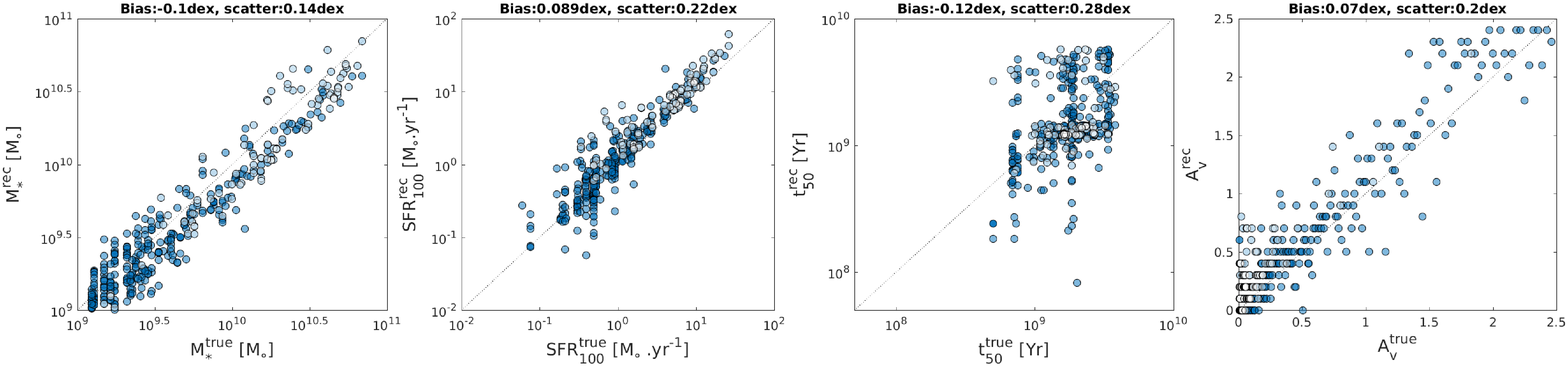}
\end{center}
\caption{\textit{Left:} Mass-weighted metallicity distribution at present epoch, binned using the BC03 Z range. Solid bars denote the distribution for the sampling in mass used in \S\ref{sec:val_sams}, while the dashed lines denote the distribution for a uniform sampling in Stellar Mass
 \textit{Right:} SED Fits to the ensemble of SAM galaxies taking the range of realistic metallicity values into account. The $t_{50}$ reconstruction has a reasonable scatter of 0.28 dex that looks poor due to the small distribution of true values. The fits are further complicated by dust, but still comparable to the ones with a single metallicity value.}
\label{fig:sam_realz}
\end{figure}

In the left panel of Figure \ref{fig:sam_realz}, we show the distributions of observed metallicities at $z\simeq 1$ in the SAMs, weighted using the dominant contribution to the total mass of the galaxy for realistic sampling in Stellar Mass ($M_*$) used in \S\ref{sec:val_sams}.  Upon examining the SEDs corresponding to a sample of galaxies of different ages generated by combining the spectra corresponding to the star formation histories in each of BC03's six metallicity bins, we find that older, more massive galaxies are more metal-rich at the observed epoch and thus show a greater deviation from the template SEDs, which currently assume $Z=0.2Z_\odot$ for the entire SFH.

In the four panels to the right of Figure \ref{fig:sam_realz} we show the reconstruction of physical quantities ($M_*$, $SFR_{100}$, $t_{50}$ and $A_v$) for the cumulative SEDs with realistic metallicities. The increased bias in the $t_{50}$ appears to be the result of poorly fitting older galaxies, which have much higher metallicities than those in the atlas. For $t_{50}^{true} \sim 3$Gyr and older, galaxies tend to have $Z > 0.4Z_\odot$ at the time of observation. This effect in addition to the narrow distribution of true $t_{50}$ causes the scatter in $t_{50}$ to appear poor even though it is comparable to the fits in \S\ref{sec:val_sams}. The recovered SFHs themselves are still representative of the true SFH of the galaxy up to a lookback time of $\sim 3Gyr$, after which the degeneracies in the $\chi^2$ surface due to the contributions from older stars, dust and differing metallicities impose larger uncertainties on the reconstruction by a factor of $\sim 1.22$.

In Figure \ref{fig:ftest_mocks}, we now focus on the results in the last columns in each mass bin. The results agree well in the high mass bin, due to roughly equal numbers of false negatives and positives. The net results in both mass bins are still acceptable, as a result of which the method is still valid even in its current simple realisation with a single-metallicity.
\vspace{12pt}

\section{Results}
\label{sec:results}
The Dense basis method of SEDfitting allows us to reconstruct the star formation histories of galaxies in a nonparametric fashion, not being restricted to the choice of a particular number or family of basis SFHs, while being able to compress the reconstructed SFHs using a small number of parameters to describe a best fit. We show the results of applying this method to our sample of CANDELS galaxies at $1<z<1.5$ and mock SAM galaxies at $z\sim 1$. 

\subsection{Going beyond 'age' and instantaneous SFR}
\label{sec:aget10}

\begin{figure}[ht!]
\begin{center}
\plotone{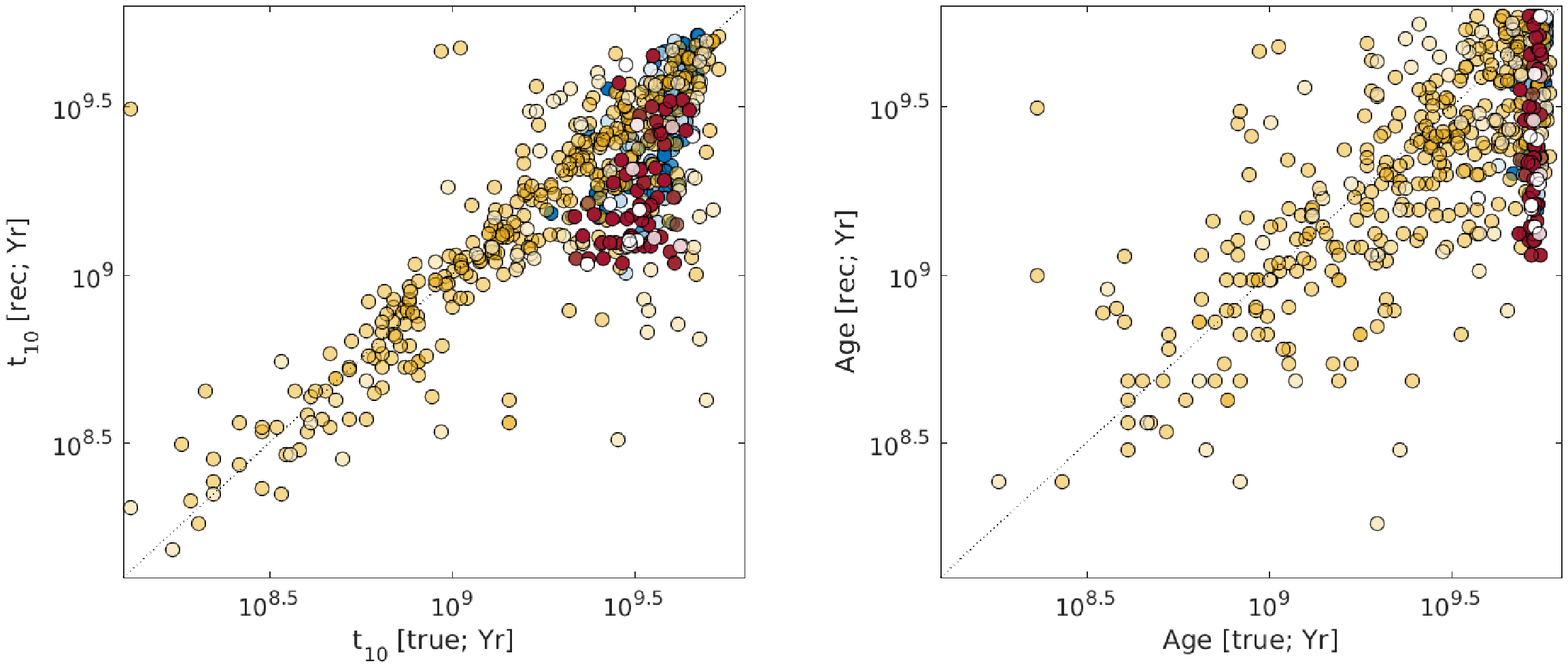}

\plotone{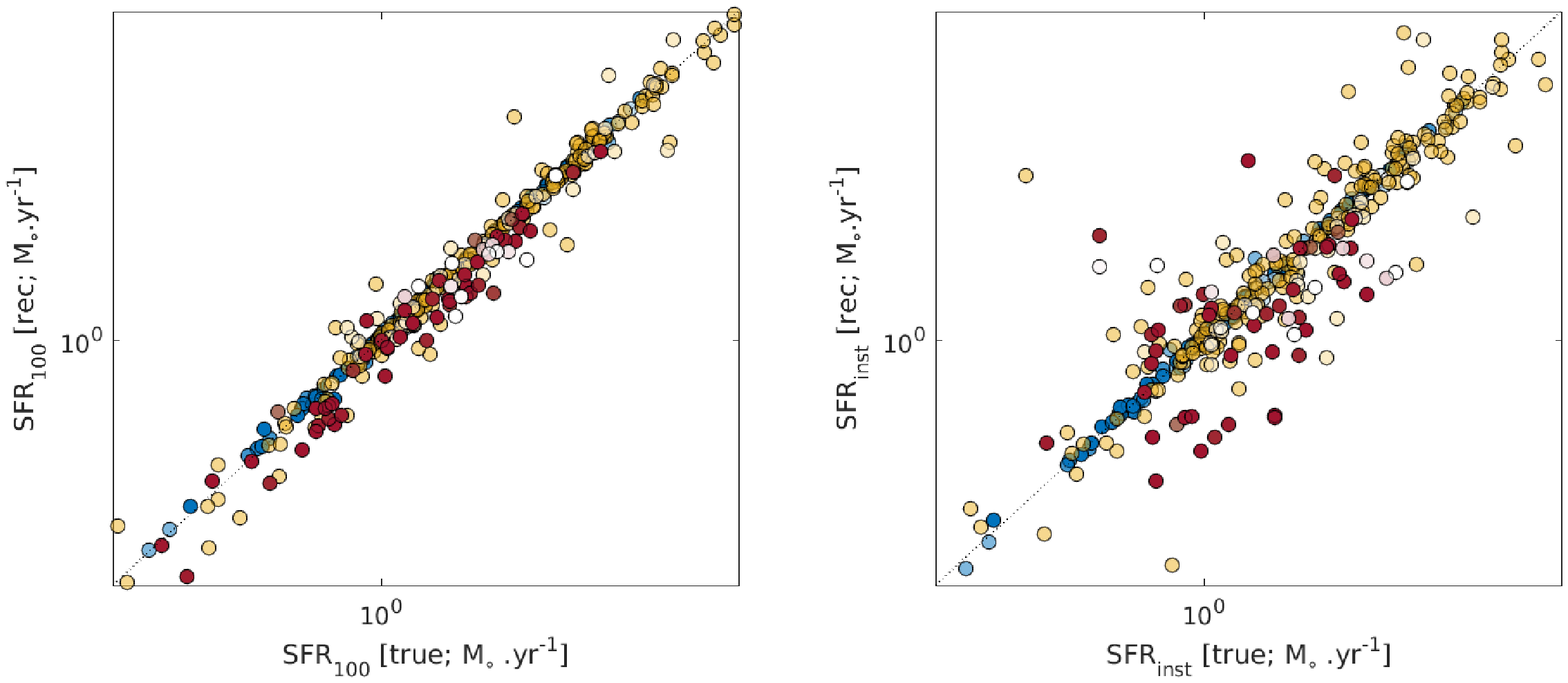}
\end{center}
\caption{\textbf{(Top:)} Plots showing the ability to extract $t_{10}$, the lookback time at which the galaxy has accumulated $10\%$ of its observed mass, is much more reliably estimated than the `Age $(\equiv t_0)$' of that galaxy with both reduced bias: -0.13dex for $t_{10}$ vs -0.19dex for Age, and reduced scatter: 0.24 dex for $t_{10}$ vs 0.31 dex for Age. \textbf{(Bottom:)} An illustration of a similar robust measure with $SFR_{100}$ showing less scatter than $SFR_{inst}$. For the $SFR_{100}$, the bias is -0.03dex and the scatter is 0.11dex. For the $SFR_{inst}$, the bias is -0.04dex and the scatter is 0.37dex. The three different colors show fits to the three different mock datasets, with blue for galaxies from the SAM, yellow for the stochastic galaxies, and red for galaxies from the hydro. simulations, using the same notation as Figure \ref{fig:threevals}.} 
\label{fig:age_vs_t10}
\end{figure}

The `Age' of a galaxy, defined as the lookback time at which the galaxy first started forming stars $(\equiv t_0)$, is not as meaningful with realistic SFHs as it used to be with simple stellar populations, which formed all their stars at a single lookback time, given by the Age \citep{tinsley1980evolution, bc03}. Realistic SFHs as seen in the SAM and the hydrodynamic simulations may maintain a small amount of star formation before ramping up to a major episode of star formation, which results in the true Age for most galaxies approaching the age of the universe. Since the SED of a galaxy is most sensitive to its largest episodes of star formation, with its sensitivity decreasing as we go back in lookback time, the 'Age' recovered through SED fitting methods is not a robust physical quantity. However, if we were to estimate the lookback time at which the galaxy  accumulated the first $10\%$ of its observed stellar mass, we estimate the lookback time at which any major star formation activity in the galaxy started. While the distributions of the Age and $t_{10}$ are similar for a given sample of galaxies, the latter is a more meaningful quantity in terms of studying galaxy growth and evolution and is more robustly estimated through SED fitting. \citep{pacifici2015importance, pacifici2016timing} This can be seen from the top panels of Figure \ref{fig:age_vs_t10}, with the right panel showing noiseless reconstructions of the Age, and the left panel showing noiseless reconstructions of $t_{10}$ for all three samples of galaxies used in \S\ref{sec:val_sams} using the same basis set and format for the plots. The latter quantity is more robust, as can be seen from the reduced bias and scatter in the estimation of $t_{10}$.

In a similar manner, due to the large amount of stochasticity that realistic SFHs show, it is more robust to estimate the Star Formation Rate (SFR) averaged over the last 100Myr in lookback time, rather than the instantaneous SFR, as shown in the bottom panels of Figure \ref{fig:age_vs_t10}. The panel on the left denotes $SFR_{100}$, which has less scatter than $SFR_{inst}$, shown on the right. It is widely appreciated that broad-band SED fitting is primarily sensitive to SFR averaged over the past 100Myr\footnote{However, when nebular emission lines are strong enough to contribute significantly to the broad-band photometry, SED fitting can probe $\sim 10$ Myr timescales.} \citep{conroy2013modeling, johnson2013measuring}, but SED fitting traditionally reports SFR$_{inst}$ in its chosen parametrization nonetheless. With rapid rises and exponential declines possible, these quantities can differ significantly, leading to the extra scatter in the bottom right panel of Figure \ref{fig:age_vs_t10}.

\subsection{The number of episodes of star formation experienced by $1<z<1.5$ CANDELS galaxies}

It is an important feature of the dense-basis method to be able to recover the number of strong episodes of star formation in a galaxy. Doing so allows us to detect recent bursts of star formation, or a period of relative quiescence between episodes of continuous star formation, with the amount of data that can be extracted depending upon the S/N. This can then be used to infer valuable information about the galaxy's evolution and merger history.

\begin{figure}[h!]
\begin{center}
\plotone{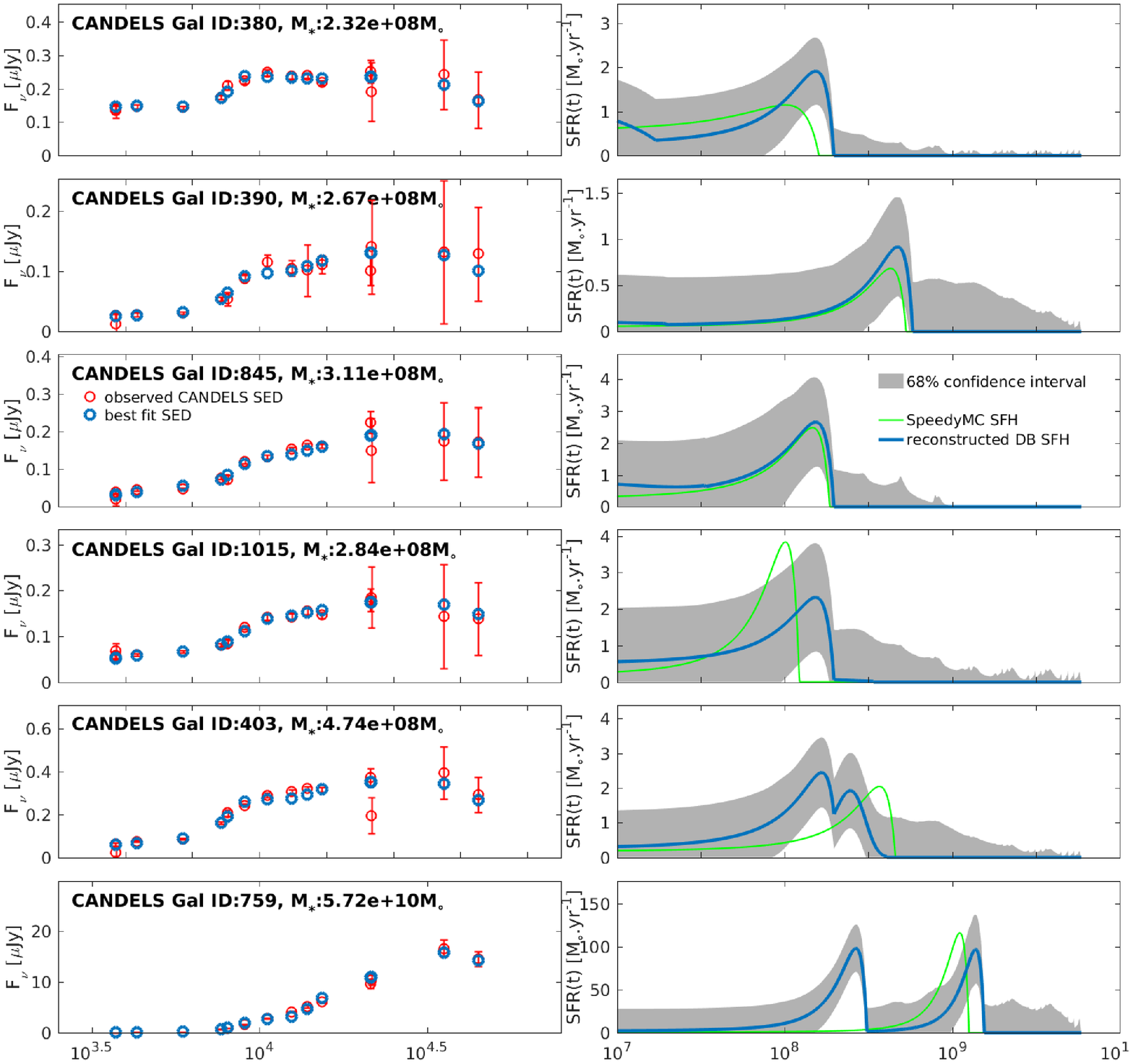}

\end{center}

\caption{ Updated to show all speedy fits. Plots showing a randomly drawn sample from the 790 CANDELS galaxies at $1<z<1.5$ used for the validation and results sections of this paper. (\textbf{Left:}) Plots show photometry from CANDELS at $1<z<1.5$ (red errorbars) and the best fit SED (blue open circles) using the Dense Basis method. (\textbf{Right:}) Plots show the single Linexp SFH fit with SpeedyMC in \citep{kurczynski2016evolution} (green line) and the Dense Basis reconstruction (blue solid line) with 68\% confidence intervals (grey shaded region) computed using the method described in \S\ref{sec:sfhuncert}. The $\chi^2$ of the fit for each of the galaxies is $12.7,7.0,30.8,8.7,40.7$ and $ 41.5$, for fits with 15 of the 17 CANDELS bands, excluding IRAC Ch.3,4 since we have not modeled for PAH emission in our atlas. The spectroscopic redshifts of the various galaxies are 1.0910, 1.1300, 1.2510, 1.3810, 1.0760 and 1.2210 respectively.}

\label{fig:candelsexamples}
\end{figure}

In this paper, we have demonstrated the use of an F-test  to detect if the addition of a second component of star formation is a statistically significant improvement to the fit. This is then used to infer the fraction of galaxies whose SFHs contain a second major episode of star formation, and was validated for the mock galaxies in the high S/N regime in \S\ref{sec:val_sams}. 
For our current sample of 1100 CANDELS GOODS-S galaxies, we can reliably fit 790 galaxies, with the remaining galaxies either having poor $\chi^2$ or with missing fluxes in multiple filters, preventing robust estimation of the SFH and its uncertainties. The F-test then determines that 134 galaxies out of the sample of 790 galaxies show a statistical improvement upon being fit with a second component, of which 117 galaxies contain a second episode of star formation. This corresponds to roughly $15\%$ of the sample, similar to the results for the mocks. Figure \ref{fig:candelsexamples} shows six examples of the procedure, showing three galaxies that were fit by a single basis SFH and three with two components.

\begin{figure}[h!]
\begin{center}
\plotone{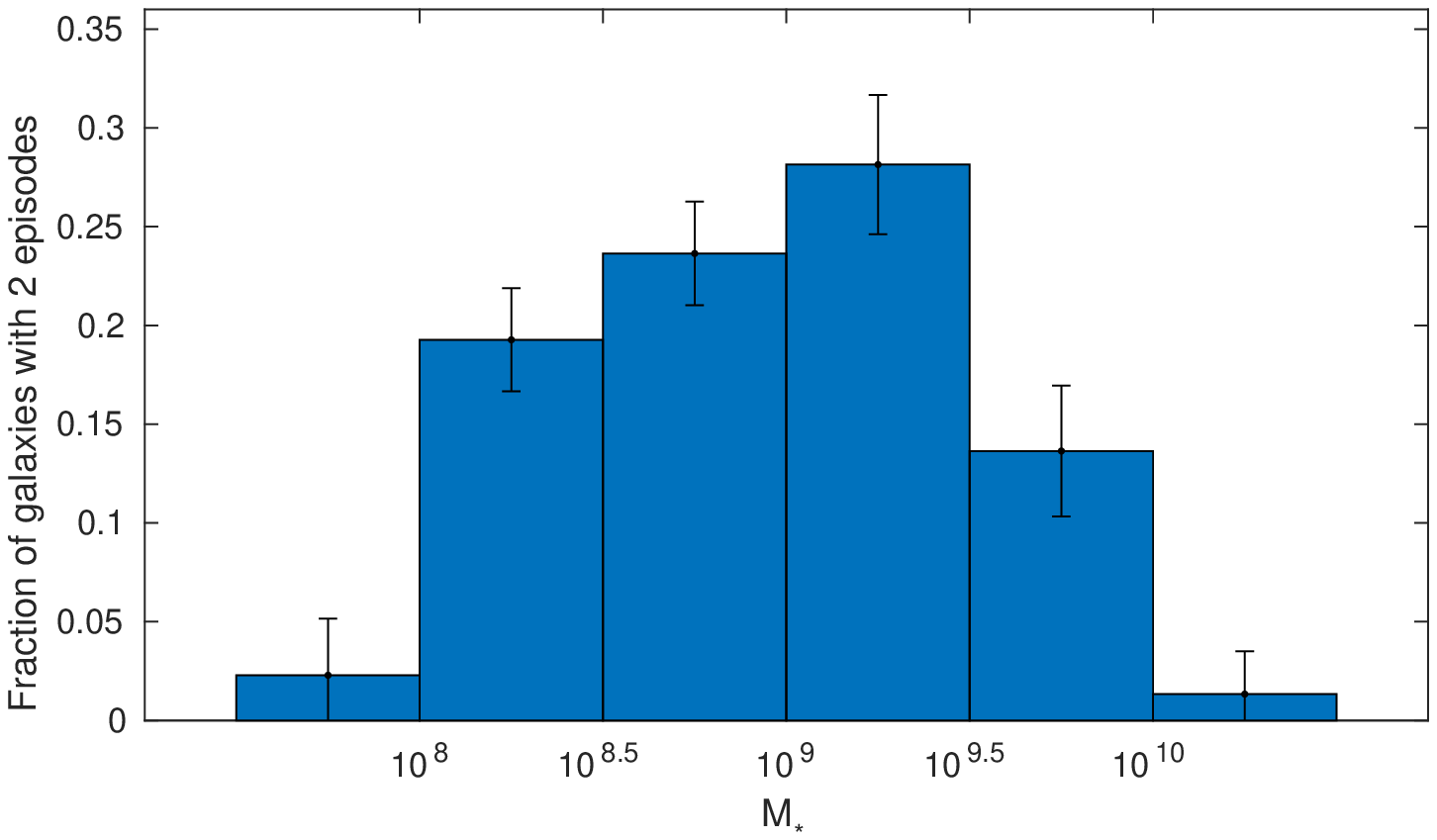}
\end{center}
\caption{We find the number of galaxies that show a statistical improvement upon being fit with a second component of star formation using the Best4 basis from \S\ref{sec:samtrain}, and then determine which of those correspond to a second episode of star formation. For the sample of $790$ CANDELS GOODS-S galaxies, $~15\%$ of the galaxies are fit with multiple episodes of star formation, with the histogram showing the distribution of the fraction of galaxies that are fit with a second episode of star formation across different ranges in stellar mass. The Poisson error bars denote the possible uncertainties due to limited sample size.}
\label{fig:ftest}
\end{figure}

Additionally, we provide a breakdown of the fraction of galaxies in each mass bin from $[10^8,10^{10}]M_\odot$ shown in Figure \ref{fig:ftest}. This figure reveals a decrease a significant decrease in the fraction of galaxies that are fit with two major episodes of star formation as the stellar mass increases above $10^{9.5}M_\odot$. As seen in \S.\ref{sec:val_sams},we expect to underestimate the fraction of 2-episode galaxies at lower masses in the CANDELS sample. Hence the increased number of 2-episode galaxies at $M_*<10^{9.5}M_\odot$ is a robust indication that 2-episode galaxies are more common at lower mass. This discrepancy between the data and simulations is intriguing.

\subsection{Constraints on timing and duration of episodes}

\begin{figure}[h!]
\begin{center}
\plotone{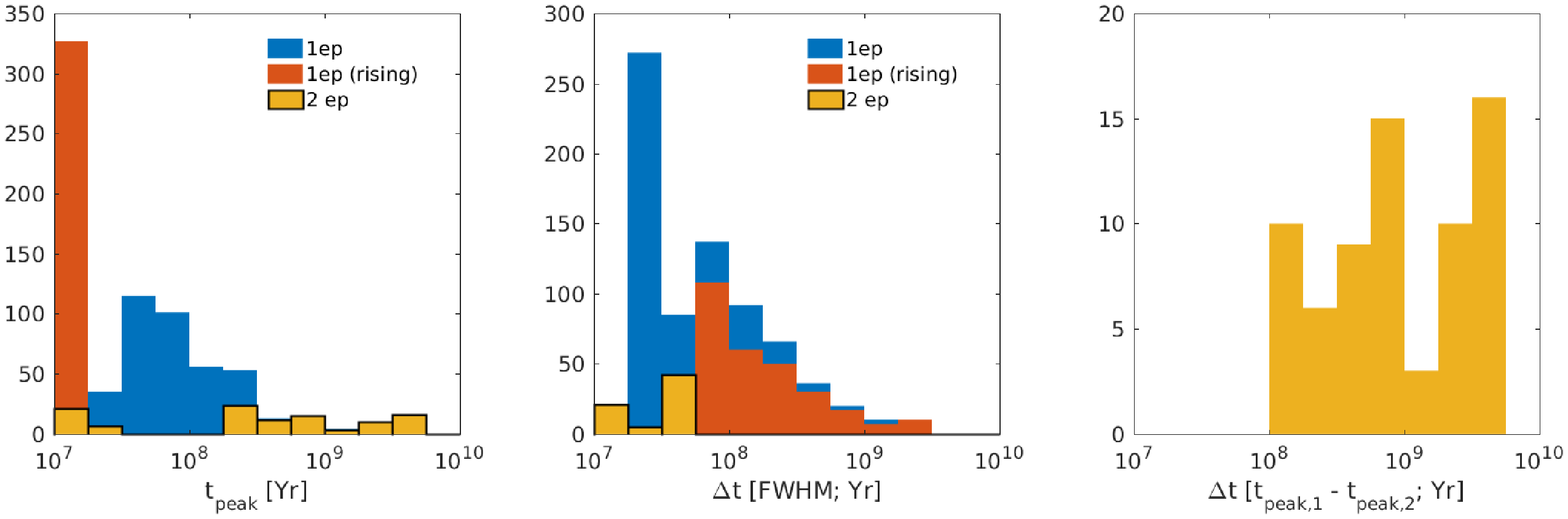}
\end{center}
\caption{\textbf{Left:} Histogram of the lookback time at which the reconstructed SFH for the sample of CANDELS galaxies peaks. The blue histogram denotes the peak times for the galaxies with a single episode of star formation while the smaller yellow histogram shows the peak times whose reconstructed SFH contains two episodes of star formation. A significant fraction of the galaxies have SFHs that are still rising at the epoch of observation, and represent $\sim 90\%$ of the first bin in the histogram, shown in red. \textbf{Middle:} Histogram of the widths of star formation episodes corresponding to the same sample obtained at the FWHM of the reconstructed SFH, with the red histogram representing the portion of the reconstructed SFHs that are still rising, with their widths truncated at the time of observation. \textbf{Right:} Histogram of the separation between the two peaks for the SFHs with two episodes of star formation.}
\label{fig:peakwidth}
\end{figure}

Using the reconstructed SFHs for our CANDELS sample, we can obtain constraints on the timing and duration of episodes of star formation. This is possible since the reconstructed SFH using our well motivated basis SFH set captures the general trend of star formation, even if the finer stochastic details are lost. For each fit, we obtain the number of episodes of star formation, the lookback time of peak star formation, and the FWHM of that episode, thus obtaining the timescale of star-formation episodes both on a galaxy-by-galaxy as well as an ensemble basis, as shown in Figure \ref{fig:peakwidth}. For $\sim 40\%$ of the galaxies, we find that the SFH is still rising at the time of observation, comparable to $ \sim 30\%$ for galaxies from the SAM and Hydrodynamic Simulation. In estimating the width of an episode of star formation, we estimate the width of an episode up to the time of observation, leading to truncated widths for the subsample of galaxies whose SFHs are still rising, shown in red in the histogram. We find that the widths for our sample are smaller by a factor of $\sim 10$ than those for the mock galaxies. This discrepancy bears further investigation, with a similar difference seen in \citet{diemer2017log}. Additionally, we can also find the interval between episodes of star formation for the galaxies that are reconstructed with two episodes of star formation.

\subsection{Statistics of $t_{10}, t_{50}, t_{90}$ and uncertainties}
\label{sec:massass}

\begin{figure}[ht!]
\begin{center}
\plotone{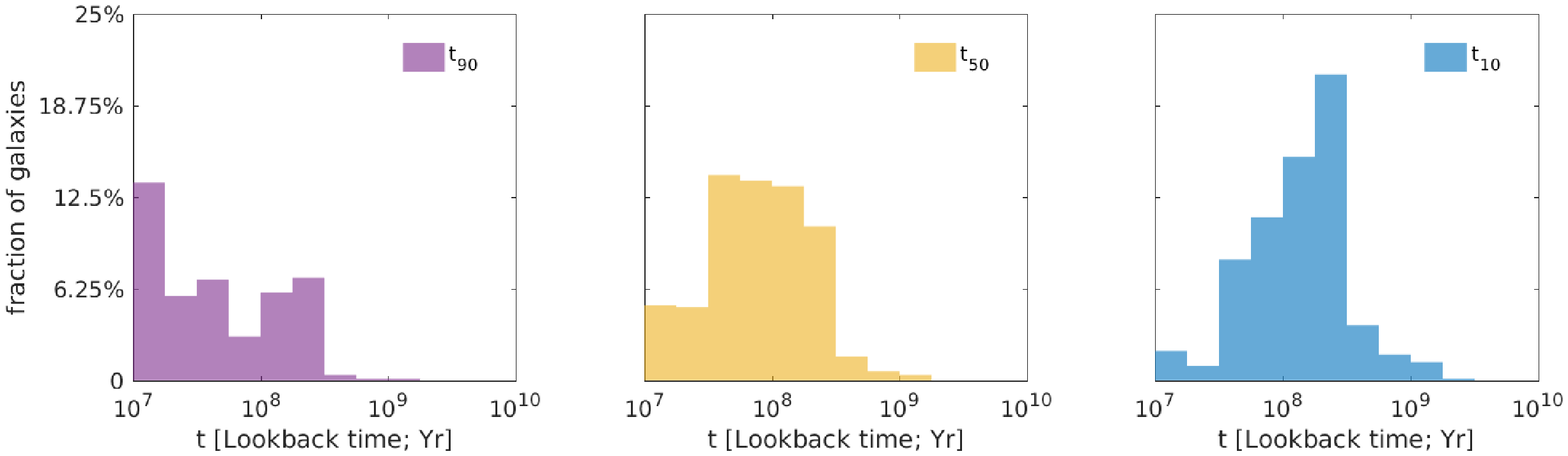}
\end{center}

\caption{Distributions of the timescales at which the galaxies in the CANDELS sample assembled $90\%$, $50\%$ and $10\%$ of their observed mass, showing the fraction of the sample vs lookback time. The purple histogram shows $t_{90}$, the yellow histogram shows $t_{50}$, and the blue histogram shows $t_{10}$. }
\label{fig:massass}
\end{figure}

The reconstructed SFHs for the CANDELS galaxies computed using the Dense Basis method are used to compute the lookback times at which the galaxy accumulates a certain fraction of its observed mass. These quantities, defined $t_x$, satisfy the equality,
\begin{equation}
\int_0^{t_x} \psi_k(t') dt' = \frac{x}{100} \int_0^{t_{obs}} \psi_k(t') dt'~~~\implies~~~M_* (t_x) = \frac{x}{100} M_* (t_{obs})
\end{equation}
Generalising $t_{10}$ from \S\ref{sec:aget10}, this lets us follow the mass assembly using the lookback times at which the ensemble of galaxies accumulated a certain fraction of its observed mass. We do this for the CANDELS sample at $1<z<1.5$ in Figure \ref{fig:massass}, providing histograms showing the overall lookback times at which the individual galaxies accumulated 10\%, 50\% and 90\% of their observed stellar mass. This allows us to infer the overall growth and evolution of galaxies at that epoch.

 \vspace{12pt}
\section{Discussion}
\label{sec:discussion}

\subsection{Biases from using single SFH parametrizations}
\label{sec:biasva}

The flexibility in the choice of SFH family used for SED fitting makes it possible to quantify the bias introduced in the estimation of physical quantities due to the choice of SFH parametrization used. We briefly list these biases at $z\sim 1$ for the six families of SFHs presented in this work, highlighting the particular families that perform best at the estimation of a particular quantity.  For the seven physical quantities $Q_i$ listed in Tables \ref{table:biasres} and \ref{table:scatterres} below, we formulate the bias and scatter as the median and standard deviation of the histogram $b(Q) = \{1 - Q_i^{rec}/Q_i^{true} \}$, which gives the scatter after taking the bias into account. This is done using physical quantities computed using the reconstructions of the SFHs of the 1200 mock galaxies from \S\ref{sec:val_sams} with a realistic mass distribution, using fits without dust or noise to highlight the bias due to the SFH parametrization. We have included the CSF, Top-Hat and Exponential  biases in Table \ref{table:biasres} below in an effort to standardize quantities in comparison to older literature, while also listing the reduced bias and scatter with the full Dense Basis method with up to two components of basis SFHs from the Linexp, Besselexp, Gaussian and lognormal families. In order to ensure a fair comparison, all families contain the same number of basis SFHs and are dense enough to converge, i.e., a denser grid on the parameter space does not change the results significantly.

\begin{table}[ht!]
\caption{Bias in the estimation of physical quantities due to different SFH parametrizations at $z\sim 1$. [No dust or noise]}
\label{table:biasres}
\begin{center}
\begin{tabular}{ c|c c c c c c c}
\hline \hline
&	 $M_*$ & SFR$_{100}$ &  SFR$_{inst}$ 	& $t_{90}$ & $t_{50}$ &	 $t_{10}$ &	 Age 	\\
\hline
CSF 		& -14\% &  5\% & 4\% & 	-24\% &  -13\% & -19\% &  -43\% \\
\hline
tophat 		 &-17\% & 2\% & 	-11\% & 	-27\% & -24\% & 	-41\% &  -59\% \\
exponential 	 &-20\% & -2\% &  -7\% & -21\% & -34\% & -55\% &	 -70\%  \\
Linexp 		& -18\% &  -1\% & -7\% &  -18\% & -28\% & -39\% & -50\%  \\
Gaussian 	 &-16\% &  -1\% & -7\% &  -20\% & -31\% & -27\% & -16\%  \\
lognormal 	 &-14\% &  2\% & -3\% & 	-16\% & -25\% & 	-33\% & 	 -26\%  \\
Besselexp 	 &-19\% &
  -1\% & -7\% &  -21\% &-34\% & -42\% &  -43\%  \\
\hline
Dense Basis 	 &-6\% &  4\% & 	1\% &  -4\% &  -4\% & -22\% & -29\%  \\
\hline
\end{tabular}
\end{center}

\end{table}

\begin{table}[ht!]
\caption{Scatter in the estimation of physical quantities due to different SFH parametrizations at $z\sim 1$. [No dust or noise] }
\label{table:scatterres}
\begin{center}
\begin{tabular}{ c|c c c c c c c}
\hline \hline
& $M_*$ & SFR$_{100}$ & SFR$_{inst}$ & $t_{90}$ & $t_{50}$ & $t_{10}$ & Age \\
\hline
CSF & 28\% &  43\% &  40\% &  36\% &  48\% &  51\% &  37\%  \\
\hline
tophat & 28\% &  50\% &  49\% &  35\% &  44\% &  46\% &  32\%  \\
exponential & 27\% &  16\% &  20\% &  34\% &  36\% &  36\% &  26\%  \\
Linexp & 29\% &  23\% &  25\% &  34\% &  38\% &  37\% &   29\%  \\
Gaussian & 24\% &  26\% &  24\% &  29\% &  31\% &  28\% &  28\%  \\
lognormal & 26\% &  16\% &  20\% &  35\% &  34\% &  31\% &  26\% \\
Besselexp & 23\% &  16\% &  18\% &  29\% & 30\% &  25\% &  27\% \\
\hline
Dense Basis & 13\% &  7\% &  10\% &  18\% & 43\% &  34\% &  20\% \\
\hline
\end{tabular}
\end{center}

\end{table}

Almost all the families tend to underestimate the stellar mass. However, the scatter in $M_*$ is generally larger than this bias except for the Top-Hat family. The scatter is even larger for SFR$_{100}$, and thus the bias doesn't significantly affect the results except at the low SFR ($SFR_{100}<10^{-1}M_\odot.yr^{-1}$) end, as seen from Figure \ref{fig:threevals}. Age has the greatest bias of all the estimated quantities, and it can be seen that it decreases when we estimate $t_{10}$, which is a more robust quantity, as we proposed in \S\ref{sec:aget10}.  About $40\%$ of the mock galaxies form $<10\%$ of their mass at $t_{lookback} > 3Gyr$. The small contribution to the observed flux from these older stars is dominated by more recent contributions, as a result of which the method does detect these older stars and underestimates the age. Since most of the mock galaxies start forming stars at $t\sim t_{bb}$, the distribution of true ages is extremely narrow and can only be underestimated, since the method does not allow $t_{age}>t_{bb}$. An artifact of this bias is also seen in $t_{10}$, although it is smaller. However, since the Dense Basis method recovers the major episodes of star formation and the bias is largely due to the distribution of the true ages, the reconstructed SFHs are robust. In a similar vein, the bias decreases in considering $t_{50}$ and even further with $t_{90}$. Age has a lower scatter than $t_{10}$, since most galaxies start forming stars at $t_{0} \sim t_{bb}$, and this creates a narrow distribution for the true Ages. For SED fitting methods that use Age, setting $Age = t_{bb}$ would lead to a bias of $-23\%$ and a scatter of $32\%$, fully competitive with any of the single families. The scatter in $t_{10}$ for the Dense Basis method is also in part due to expanding to a larger parameter space, which yields a smaller bias at the expense of increased scatter regulated by the F-test. The Top-Hat, exponential and Linexp parametrizations have a large bias in age, and should be accounted for in comparisons of ages in the literature. $t_{90}$ is the most robust of the mass-assembly times, with the Linexp and lognormal families performing best in its estimation. The Dense Basis method offers the least scatter in M$_*$, SFR, $t_{90}$ and is nearly unbiased in these quantities, as well as $t_{50}$.

\subsection{Comparison with other methods}
\label{sec:compareparam}

The Dense Basis method offers an SED fitting approach that minimizes the bias and scatter introduced due to traditional SFH parametrizations. In this section, we consider comparisons with existing methods of SFH reconstruction. MOPED \citep{moped} fixes logarithmic time bins and finds the SFR in each bin with a piecewise constant SFH using fitting with data compression, giving more weight to those pixels in the spectrum that carry most information about a given parameter. VESPA \citep{vespa} adaptively bins the lookback time, i.e., the $t_i$ in Eq.11, provided there are enough free parameters to avoid overfitting. Dye's (2008) method adopts a similar approach with photometry, but uses regularization in order to the make the SFR in each bin positive, which might bias the likelihood surface and is computationally more expensive. None of these methods reconstructs smooth SFHs; the fits do not provide us with SFHs that allow us to analyse the number of episodes of star formation or to analyse the peak times and widths of star formation episodes. The method introduced here uses a physically motivated functional form of SFHs that requires a smaller number of free parameters to fit the SFH, thus obtaining smooth SFHs with multiple components through photometric SED fitting, comparable to what was previously accessible with spectroscopy or CMD reconstruction \citep{weisz2011acs}. Another advantage is the ability to use real SEDs to test functional forms for a best match against star-formation mechanisms at a given redshift. The usage of well-motivated parametrised functional forms instead of time bins with variable heights allows us to obtain a smooth reconstruction of the SFH with a smaller number of parameters without the need for regularization, since the basis SFHs are smooth and positive definite.

\subsection{Possible extensions of the Dense Basis method}
\label{sec:future}

In addition to the two parameter families described in \S\ref{sec:formulation}, it is possible to extend the approach to a larger parameter space by using families of curves including the 4-parameter families described in \citet{simha} and the Exponential+ power law \citep{behroozi},
\begin{equation}
f(t,t_0,\tau,\alpha) = \Theta(t-t_0) ((t-t_0)/\tau)^\alpha e^{-(t-t_0)/\tau}
\end{equation}
where $\tau, \alpha \in \mathbb{R}_+$, and $t_0$ indexes the time at which star formation begins. 

Currently, however, we restrict our attention to the two parameter families since we also consider combinations of curves from these families, which let us model a much more versatile set of trajectories in SFH space.

An advantage of our method is that it will recover only as many SFH basis components as are needed to produce a good fit to the SED, thus enabling us to extend the procedure to reconstructing metallicity histories, and to use multiple dust extinction models. It is also possible to extend the code to additional SPS models, which is naturally incorporated with the Conroy FSPS models \citep{conroy2010propagation} that contain the BaSTI and Padova isochrone sets. Model dependency due to the choice of tracks and IMF is also an issue that could be incorporated into future versions, which will have more data available that can be used to address degeneracies between different sets of isochrone synthesis models, stellar evolution tracks, and IMF choices.

Additionally, the superposition of `stochastic' bursts on top of these smooth functional forms has been better shown to reproduce the observed spectroscopic properties of individual galaxies \citep{kauffmann2003stellar, brinchmann2004physical}. This can be explored in future applications of the dense basis method to spectroscopic data, using realistic stochastic SFHs as in the approach of \citep{pacifici2015importance}, or the theoretical stochastic SFHs from \citep{kelson}.

The current formulation is frequentist, and the training and validation produce parameter uncertainty estimates consistent with this approach. A Bayesian formulation of the method is certainly possible, but since the priors on SFHs are poorly known, significant care would be required.

\subsection{Handling Big Data}

A large amount of data will be generated from the upcoming generation of surveys including LSST \citep{lsst}, HETDEX/SHELA \citep{hetdex} and J-PAS \citep{jpas}, which will yield a mixture of broad-band photometry and spectroscopy for $N\sim\mathcal{O}(10^8)$ galaxies. 

Methods for analysing these galaxies using SED Fitting techniques need to be both computationally efficient as well as capable of handling and storing large volumes of data in a memory-efficient manner.

The dense basis method was designed with these two requirements in mind. It takes $\mathcal{O}(0.1)s$ for a single run on a 2.9 GHz laptop, albeit with large memory requirements for storing the 2-component basis, which runs to $\mathcal{O}(200Gb)$ with 18 values of $\tau$ and $99$ values of $t_0$. After the initial generation of the atlas, the fits themselves can be stored simply by saving the index of the best-fit SED and the normalization for each component, leading to efficient storage of the fits as $(N_{component}*3)$ coefficients for each reconstructed SFH.

\subsection{Broader Data science applications}

This method can be used to solve problems of the general type
\begin{equation} \label{eq:big_eqn}
\mathbf{d} = \sum_i \int_t \mathbf{m}_i(t) dt \equiv \sum_i \sum_j a_j \mathbf{m}_{ij} 
\end{equation}
where \textbf{d} represents a vector of observables, i.e., galaxy SEDs in the current work, and the functionals $m_i$ represent possible SFHs. The index $i$ sums over basis functions and $j$ refers to multiple photometric bands. We adopt functionals that can be shifted by varying $t_0$ and scaled by varying $\tau$ because this is reasonable for the underlying physics of star formation. This is not a requirement for solving Eq (\ref{eq:big_eqn}) and additional constraints upon the functionals will depend upon the problem being considered. Upon generalization, this formulation is particularly useful for the class of problems where constrained observed data is used to recover quantities in an otherwise inaccessible parameter space, such as single-epoch observations of historical processes. In the absence of a known analytic mapping from the parameter space $\{ m_i(t) \}$ to the space of observables $\{ d_j\}$ and the lack of a definite correlation between the goodness of estimation in these two spaces, traditional methods like Monte-Carlo estimation through the parameter space need not lead to accurate estimation of the $m_i$, since a good fit need not correspond to an accurate reconstruction of the functional. Methods like Principal Component analysis may be used in the parameter space, but the principal components do not always correspond to physical representations of the observables. Such situations can frequently arise due to the presence of noise and degeneracies between different parameters that affect the observables.

In such cases, it is possible to apply the training method described in the current work, based on pruning a training atlas from a large space of informed estimates from empirical observations and statistical motivations, leading to an oversampled nonorthogonal `Dense Basis'. This lets us perform any subsequent fitting to the data in a subset of parameter space where the correspondence between the goodness-of-fit and goodness-of-reconstruction exists and is well defined. Since the functionals in the parameter space are well motivated, they do not span the space of all observables and are robust to noise that would correspond to `unphysical' results. In the current framing, the method is readily applicable to timeseries problems, where the observables are integrated quantities depending on the overall nature of the timeseries.

In an expanding arsenal of data-science tools, the Dense Basis method provides a convenient formalism to solve the above class of problems in a tractable manner, and to train and implement a solution finding method. The advantages of using this method  include not having the constraints of regularization imposed by matrix inversion methods or suffering from the lack of correlation between observables and principal vectors in solution space that techniques like PCA exhibit, while also being robust to noise.  
\vspace{12pt}

\section{Conclusions}
\label{sec:conclusions}

The standard assumption of a simple parametric form for galaxy Star Formation Histories (SFHs) during Spectral Energy Distribution (SED) fitting biases estimations of physical quantities and underestimates their true uncertainties. In this paper, we introduce the Dense Basis method, which offers a general approach when a vector of observed data points \textbf{d} can be modelled as a sum of positive-definite, continuous functionals \textbf{m}$_i$ obeying \textbf{d} $= \sum_i \int_t $\textbf{m}$_i (t) dt$. Here we apply it to the case where \textbf{d} represents a galaxy SED and the functionals are possible SFHs. 

We train the method using SFHs from mock catalogs at $z\sim 1$ from three different sources: a Semi Analytic Model (SAM), meshless hydrodynamic simulations, and stochastic realizations. We do this to ensure that the method can successfully reconstruct a wide variety of SFHs allowing us to relax the assumption that our training SFHs are prefectly representative of the true SFHs of galaxies at that epoch. The training step allows us to compare the goodness-of-fit in SED space to the goodness-of-reconstruction in SFH space. We use this comparison to eliminate SFH families that provide poor or biased reconstructions, leading us to drop the Top-Hat and Exponential families from our basis, while keeping the Linexp, Besselexp, Gaussian and Lognormal families.

A basis consisting of these four families and their combinations is then used to apply the Dense Basis method to the broad-band CANDELS photometry of a sample of galaxies at $1<z<1.5$. The method allows us to accurately estimate physical quantities of interest that explicitly depend on the SFH, notably Stellar Mass (M$_*$), and SFR$_{100}$, which we note is more robust than SFR$_{inst}$ and dust attenuation. The method also allows us to estimate previously inaccessible quantities, including the number and duration of star formation episodes in a galaxy's past, and the lookback times at which the galaxy accumulates $10,50,90\%$ of its observed mass, which are more robust quantities than the \textit{Age} of a galaxy, and allow us to track the galaxy's growth and evolution as a function of lookback time. The current frequentist implementation of the method allows us to estimate confidence intervals for these quantities. We quantify the bias and scatter in these quantities due to various SFH parametrizations including the traditional parametrizations of constant and exponentially declining SFHs. 

The method can be expected to have broad data science applications, and can be scaled and applied to high S/N spectrophotometry from upcoming surveys across all redshift ranges to reconstruct the SFHs of individual galaxies, as well as to infer the growth and evolution of the ensemble of galaxies at various epochs. 
\vspace{12pt}

\section{Acknowledgements}

The authors would like to thank the anonymous referee, as well as Viviana Acquaviva, Robin Ciardullo, Caryl Gronwall, Alex Hagen, Boris Leistedt, Regina Liu, Camilla Pacifici, David Spergel, Min-Ge Xie and Greg Ziemann for their insightful comments and suggestions, Rachel Somerville for providing the SAM SEDs and SFHs, Peter Kurczynski for providing the SpeedyMC results from Kurczynski et al. (2016), and Romeel Dave for providing the MUFASA SFHs. The authors acknowledge support from Rutgers University. This material is based upon work supported by the National Science Foundation under Grant No. AST-1055919. Support for Program number HST-AR-14564.001-A was provided by NASA through a grant from the Space Telescope Science Institute, which is operated by the Association of Universities for Research in Astronomy, Incorporated, under NASA contract NAS5-26555.
\vspace{12pt}

\appendix
\vspace{12pt}

\section{Consistency across filter curves}

It is possible to perform fits to the mock galaxies observed at different redshifts and ensure that the reconstructed SEDs yield physical quantities that are robust independent of the redshift. This analysis can be extended to determine the redshift range across which a given atlas is robust, since the amount of information contained within the filters changes with redshift. 

Since we restrict our mock dataset to $z=1$ and the observed dataset to $1<z<1.5$ in current work, we perform this consistency check fitting the same mock galaxies whose rest frame spectra are computed considering them to be at $z=1$ and $z=2$. We perform Dense Basis fitting on the galaxies, and compare the derived quantities $t_{10},t_{50},t_{90}$ and find that the estimation of these quantities remains robust within uncertainties.
\vspace{12pt}

\section{Dot-product SED fitting as a computational speedup}

We present an additional approach to finding the optimal reconstruction given an atlas of SEDs using a non-orthogonal dot-product, i.e., a projection product, that might prove to be a useful computational speedup for dealing with large datasets.

Since the projection product is done in a non-orthogonal basis, reconstruction of the original vector using the dot-product coefficients is more involved than the procedure in the case of the inner product in an orthogonal space. Various methods have been tested for this reconstruction, including iteratively refitting the residuals as long as they remain above the noise level, constructing a reduced orthogonal space by projecting out components of vectors along a principal component, and constructing an expanded basis of linear combinations of the basis functionals. This method is expected to operate on the timescale of $\mathcal{O}(N\times M)$ operations, where N is the size of the basis and M is the number of bands of the photometry in consideration. 

The best-fit is estimated through a non-orthogonal equivalent of a dot product in the photometric vector space, through a mapping given by,
\begin{equation}
\phi(F_{i_1 j}, F_{ij}^{obs}) = \frac{\sum_j F_{i_1 j}.F_{ij}^{obs}}{||F_{i_1 j}||~||F_{ij}^{obs}||} = a_{i_1}
\end{equation}
where $\phi$ is a mapping such that $\phi: \mathbb{R}^{Nfilt,+} \times \mathbb{R}^{Nfilt,+} \to \mathbb{R}[0,1]$. 

For an equivalent orthogonal basis, the dot product coefficient is given by the same mapping, with an additional constraint imposed due to orthogonality, which is,

\begin{equation}
\phi(F_{i_1 j},F_{i_2 j}) = 0
\end{equation}
which allows us to reconstruct the original vector simply using 
\begin{equation}
F_{ij}^{obs} = \sum_i^{N_{basis} \leq N_{bands}} a_i F_{ij}
\end{equation}
However, in the absence of orthogonality, we turn to more involved methods of reconstruction, bounded by both computational costs and error margins on the photometry, which could lead to overfitting if not accounted for.

Other factors held constant, the coefficient of the photometric dot-product indicates the projection of the true SFH of the galaxy on to the basis SFH. Therefore, without any degeneracies in the basis SEDs, a higher coefficient would mean that the SFH is closer to the true SFH of the galaxy, with $a_i=1$ denoting a perfect match with basis vector $i$. The procedure returns similar results to the $\chi^2$ fitting procedure described in \S2.4, with a slight computational speedup requiring $\sim 1/3^{rd}$ of the time for fitting an SED, which might be helpful in fitting large datasets of SEDs from upcoming surveys.
\vspace{12pt}

\section{Alternative methods of defining goodness-of-reconstruction:}
\label{sec:r2etc}

Given that the SFHs are not a directly measurable quantity, care must be taken in comparing the reconstructed SFHs to the true ones, accounting for the unequal sensitivity of the SEDs to the same interval of time at different epochs, as well as the large amount of stochasticity present in the simulated SFHs (SAM, Hydro., Stochastic). We outline some of the methods viable for this as alternatives to be considered in other applications of the Dense Basis method. These statistics, while useful for comparing how well a given reconstruction approximates the true SFH, are significantly affected by stochasticity. Since we are only interested in the relative performance of the families of curves in current work, we choose to compare the goodness-of-reconstruction to that of a polynomial fit with the same number of degrees of freedom as the parametrizations under consideration.

\begin{enumerate}

\item \textbf{$R^2$ and $R^2$ adjusted:}

The coefficient of determination is among the simplest ways to compare two sets of points, comparing how well the reconstructed SFH approximates the true one. This gives the first indication of the fact that some families of SFHs may be more useful than others for a given dataset at SFH reconstruction.

The $R^2$ statistic is given by,
\begin{equation}
R^2 = 1 - \frac{\sum_t (\psi_{rec}(\log(t)) - \psi_{true}(log(t)))^2}{\sum_t (\psi_{true} (\log(t)) - \langle \psi_{true} (\log(t)) \rangle)^2}
\end{equation}
which quantifies the amount of variance explained by the fit. Since the stochasticity of the different mocks differs, the median $R^2$ for fits to the three datasets can vary widely, with the SAM galaxies doing the best and the MUFASA galaxies doing the worst. It is possible to adjust this by smoothing the true SFHs using a nonparametric method until they all exhibit an equal level of stochasticity, or simply by rebinning the SFH with a time interval of the order of the least stochastic sample. Another improvement, as implemented in the current work, is to compare the $R^2$ of the reconstruction with a reference $R^2$ with the same number of degrees of freedom, such as a fit using a polynomial. 

\item \textbf{The Pearson correlation coefficient} ($\rho_p$)

Since the standard implementation of $R^2$ as a goodness-of-reconstruction metric fails to account for the different amounts of stochasticity present in the different mock datasets, we consider the Pearson correlation coefficient, which accounts for the inherent stochasticity of an SFH through a normalization. As an alternative to the previous method, we can present the results for the training step as likelihood vs the Pearson correlation coefficient, written as
\begin{equation}
\rho_{true,rec} = \frac{cov(true,rec)}{\sigma_{true}\sigma_{rec}} =  \frac{\sum_i (x_i-\bar{x})(y_i-\bar{y})}{\sqrt{\sum_i(x_i-\bar{x})^2}\sqrt{\sum_i (y_i-\bar{y})^2}}
\end{equation} 
since this could better provide an estimate of the goodness-of-reconstruction for highly stochastic SFHs without the need for an additional $R^2$ adjustment step. However, the coefficient in this form assumes Gaussian statistics which are not always applicable for our datasets.

\item \textbf{Spearman's correlation coefficient} ($\rho_s$)

Pearson's correlation coefficient assumes a Gaussian distribution of noise around a linear relation, and finds the degree of correlation around it. However, the relation we seek, to compare two time-ordered sets of curves, needs to be more robust. Therefore we considered the Spearman coefficient, which compares two monotonic functions using ranks in order to find the degree of correlation between them. For distinct ranks, the coefficient is given by $r_s = 1-6\sum d_i^2/n(n^2-1)$, where $d_i$ is the difference between the two ranks of each observation. This, however does not work very well at describing the fit for young galaxies, where a significant fraction of the two star formation histories is tied at the same rank due to long periods of vanishing SFR at early times. 

\item \textbf{MISE}

The mean integrated square error given by $MISE = \sum_t \psi_{rec} (log(t)) - \psi_{true} (log(t))$ also provides a method to quantify the goodness-of-reconstruction. However, it does have a well defined range to compare different quantities, and provides no accounting for the varying amounts of stochasticity of the different datasets. The concept of minimum distance estimation that this method implements can also be generalized to the Kolmogorov-Smirnov statistic, which depends on the maximum absolute difference between the true and estimated cumulative SFHs, but does not provide sufficiently sensitive results to make a distinction between the different families using a correspondence between the statistic and the goodness-of-fit.

\end{enumerate}
\vspace{12pt}

\section{Robustness to noise}
\label{sec:noise_robust}

\begin{figure}[ht!]
\plotone{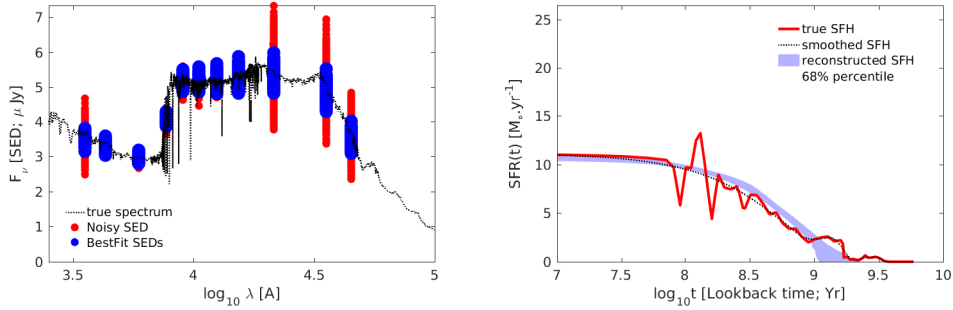}
\caption{\textbf{Left:} Fits using the Dense Basis method (blue circles) to 1000 noisy realisations (red circles) of the photometry corresponding to a galaxy with true spectrum (black dotted line) showing the robustness of the method to noise that corresponds to unphysical regions of the SFH space. Spectra are shown without nebular emission for clarity. \textbf{Right:} The pointwise 68\% intervals of the reconstructed SFHs for each noisy realisation (blue shaded region) compared to the true SFH (red solid line) showing that the reconstructions are also largely robust to the noise.}
\label{fig:noiseresist}
\end{figure}

\begin{figure}[ht!]
\plotone{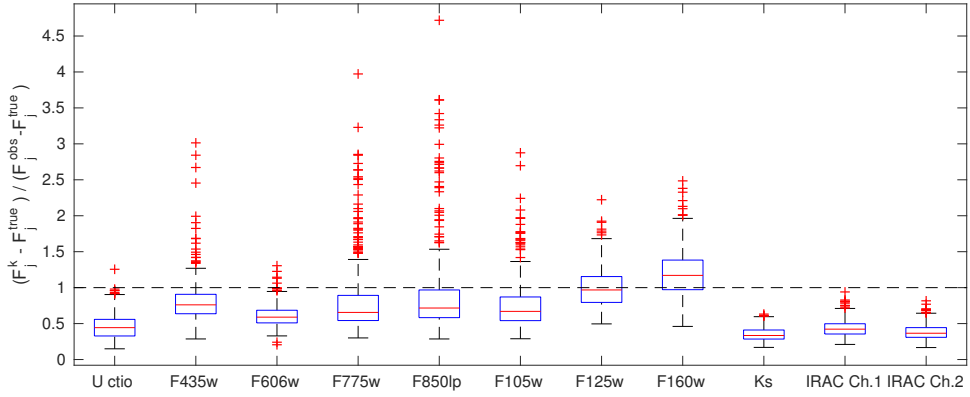}
\caption{Boxplots showing the results of fitting the ensemble of mocks with multiple noisy realisations and comparing the residuals of the fits to the noise.}
\label{fig:bandboxplots}
\end{figure}

The Dense Basis method performs SED fitting using an atlas of SEDs corresponding to well motivated basis SFHs that satisfy the conditions in \S\ref{sec:formulation}. Although the mapping from SFH space to the observed photometry is theoretically bijective, an SED at a given noise level for a given set of photometric bands is degenerate in SFH space to the extent that all the SFHs that produce the same SED within error limits are an acceptable fit. Our formulation then ensures that the basis is effectively dense in SFH space, allowing us to reconstruct the overall trend of star formation even if it doesn't capture the finer stochastic details. However, even though the basis is effectively dense in SFH space, it is not dense in SED space, since a large region of the photometry space is not accessible through any physically motivated SFH. This allows our method to ignore all noise that is `unphysical' while performing the SED-Fitting step. Even though this yields worse $\chi^2$ and the noise biases the reconstruction to an extent documented in Figure \ref{fig:threevals}, it does not overfit the SED by fitting for any noise that does not correspond to a physically motivated SFH. This allows our method to be robust to a large fraction of the noise, as is seen in Figure \ref{fig:noiseresist}, where we show an example of 1000 noisy realisations to a SAM spectrum in red and the corresponding reconstruction in blue, which successfully ignores major outliers in fitting the SED. Extended to the entire ensemble of 1200 mock galaxies, we find that the ratio of the residuals to the noise is $\sim 45\%$, with a standard deviation of $\sim 9\%$, i.e.,
\begin{equation}
\frac{|F_j^k - F_j^{true}|}{|F_j^{obs} - F_j^{true}|} \approx \mathcal{N}(0.45,0.09)
\end{equation}
The decomposition of this quantity into the sensitivity to noise in individual bands in Figure \ref{fig:bandboxplots} shows that the F160w is the most sensitive to noise, with the method being remarkably robust to the noise in the ground-based bands. The maximum deviation due to noise is computed and found to be in the bounding bands (u ctio and IRAC 4.5$\mu$m). This is expected, since the endpoints are the most unconstrained in the fitting process.

\vspace{12pt}

\section{Examining the $\chi^2_{SED}-\chi^2_{SFH}$ correspondence for individual galaxies}
\label{sec:chi2surfindgals}

\begin{figure}[ht!]
\begin{center}
\plotone{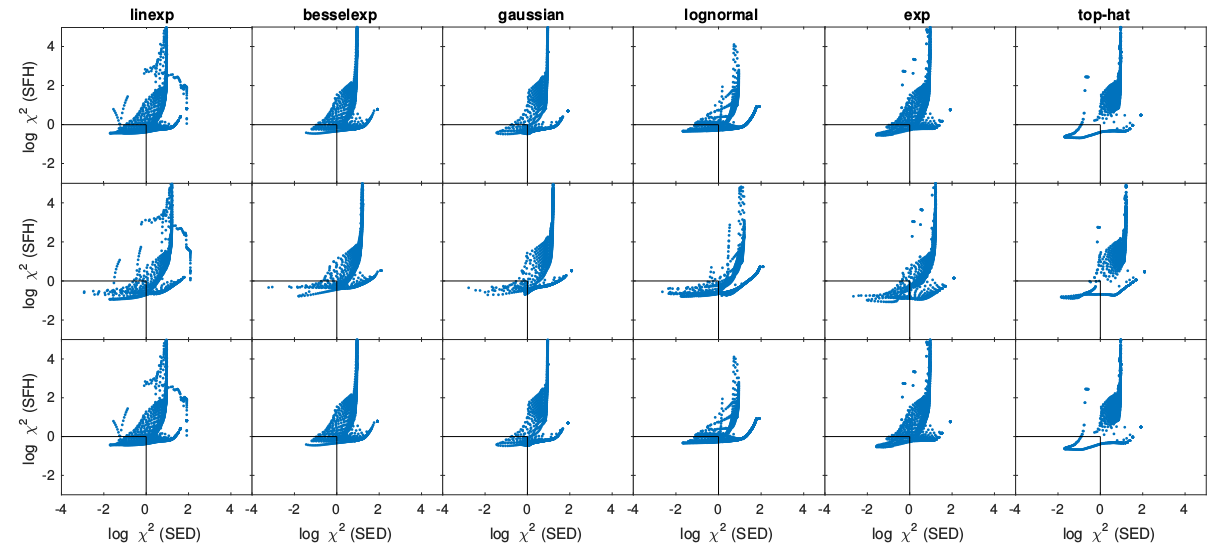}
\plotone{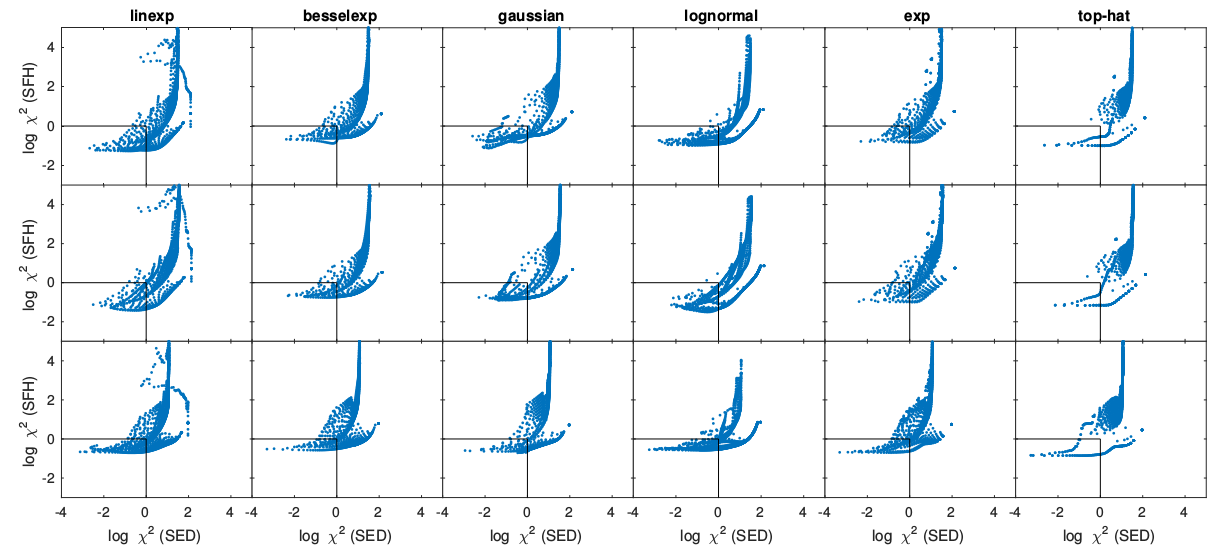}
\plotone{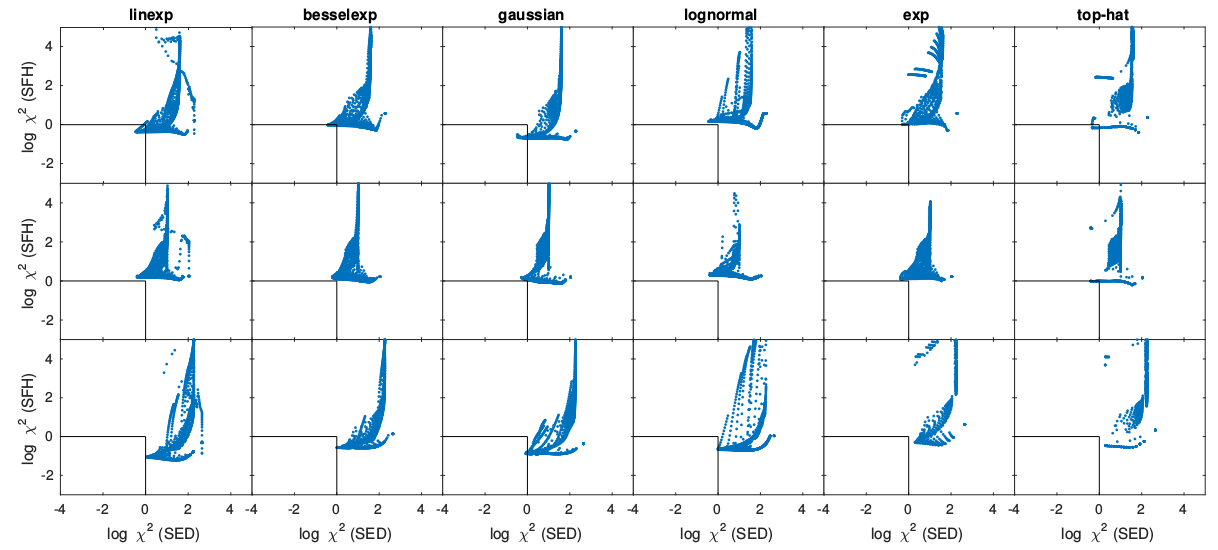}
\end{center}
\caption{Plot of the correspondence between $\chi^2_{SED}$ and $\chi^2_{SFH}$ for three randomly selected galaxies from each mock dataset, using all six families as the basis. The top three rows are galaxies drawn from Semi-Analytic Models (galaxy id = 204,278,243), the middle three rows from the Hydrodynamical simulations (galaxy id = 270,5,228) and the last three rows from stochastic realisations (galaxy id = 74,109,60).}
\label{fig:chi2surf_examples}
\end{figure}

We examine the correspondence between $\chi^2_{SED}$ and $\chi^2_{SFH}$ for individual galaxies in greater detail. We provide examples for three randomly chosen galaxies from each mock catalog as examples in Figure \ref{fig:chi2surf_examples}, and discuss possible biases and how they should be minimised. We compute $\chi^2_{SFH}$ as follows:
\begin{equation}
\chi^2_{SFH}  = \int d(\log t) \frac{(\sum_{k=1}^{N_F} \epsilon_k \psi_k(t) -  \psi_{true}(t) )^2}{\sigma_{SFH}(t)^2}
\end{equation}
where the index k sums over the entire basis of SFHs as above, and $\epsilon_k$ denotes the stellar mass normalisation. The $\sigma_{SFH}(t)$ denote symmetric pointwise uncertainties computed through the procedure described in \S.\ref{sec:sfhuncert} for each family being tested. Since the code is implemented over a grid, the integral over time is effectively a sum over discrete time intervals as described in \S.\ref{sec:formulation}. The $R^2$ statistic computes the accuracy of reconstruction and is better for training the basis families, since it does not reward SFH families that yield larger uncertainties, as opposed to $\chi^2_{SFH}$, which does so. However, we can use this statistic to observe the correspondence between fits in SED space and reconstructions in SFH space on a galaxy by galaxy basis, and study sources of biases in the reconstruction.

We encounter two types of biases in the $\chi^2$ plots, summarised as follows:
\begin{itemize}
\item \textbf{Degenerate $\chi^2_{SED}$:} If, in addition to the correspondence, some good fits to the SED ($\chi^2_{SED}/DoF < 1$) correspond to bad reconstructions of the SFH ($\chi^2_{SFH}/DoF > 1$), the SFH reconstruction may be biased. However, these are often removed as outliers in the procedure used to compute uncertainties, as described in \S.\ref{sec:sfhuncert}.
\item \textbf{Sub-optimal $\chi^2_{SFH}$:} The best fit to the SED corresponds to a significantly worse reconstruction than the best possible reconstruction of the SFH with that basis. However, like the true SFH, the best possible reconstruction is generally within our reported uncertainties around the best-fit determined via $\chi^2_{SED}$.
\end{itemize}

\begin{figure}[ht!]
\begin{center}
\plotone{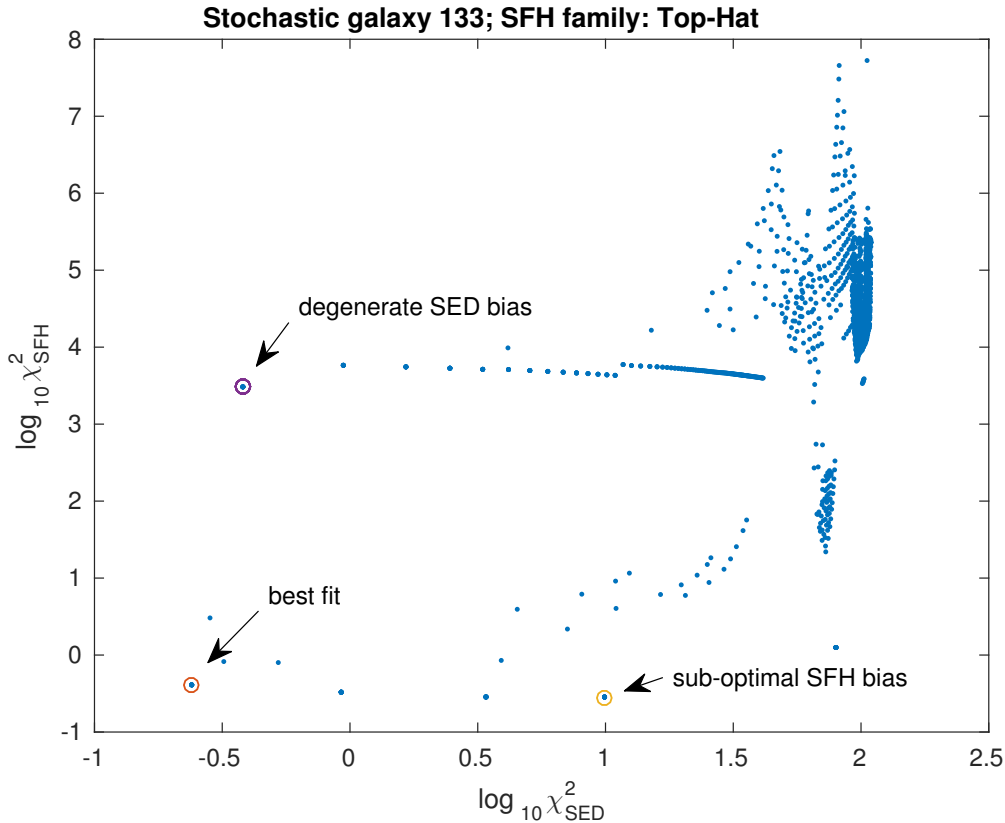}
\end{center}
\caption{Plot of the correspondence between $\chi^2_{SED}$ and $\chi^2_{SFH}$ for an individual galaxy with the Top-Hat family of SFHs, computed using noiseless fits to the SED. This illustrates the biases that can occur in SFH reconstruction through SED fitting, which we try to minimise through the training procedure.}
\label{fig:chi2surf_bias}
\end{figure}

For the first point, we quantify the two kinds of biases using the $\chi^2$ surface generated for each galaxy in the ensemble of 1200 galaxies using each SFH family. An example of the two kinds of bias is shown in Figure \ref{fig:chi2surf_bias}, showing the $\chi^2_{SED} -\chi^2_{SFH}$ plot for a single galaxy with a single SFH family. Since there is a certain amount of degeneracy introduced in SED fitting due to noise, we consider the set of all good fits ($\chi^2_{SED}/DoF<1$) instead of the best fit $min(\chi^2_{SED}/DoF)$. As shown in Figure \ref{fig:chi2surf_examples}, we see that there is generally a good correspondence between $\chi^2_{SED}$ and $\chi^2_{SFH}$ in the regime of good fits. We then find the families that minimise the two types of biases in SFH reconstruction.

We quantify the degenerate $\chi^2_{SED}$ bias by examining the histogram of the set $S = \{\chi^2_{SFH,i} ~|~ \chi^2_{SED,i} < 1\}$. Since this is the set of good fits, we then say that a galaxy has a type 1 bias if it has multiple peaks in this histogram, separated by a minimum distance of 1 dex in $\chi^2_{SFH}$. This is the more common type of bias, and the probability that it will bias the fit towards a poorer reconstruction depends on the ratio of the areas under the two peaks. We show the number of occurrences of this type of bias for each family in Table \ref{table:gofgor}, finding that the exponential and CSF families have the highest occurrence of this behaviour. While these biases are more common than sub-optimal $\chi^2_{SFH}$ biases, they only indicate the possibility of a bias due to noise, and are usually much milder than the example shown.

\begin{table}[ht!]
\caption{Comparison of the Goodness of Fit to the Goodness Of Reconstruction for different samples of mock SFHs}
\label{table:gofgor}
\begin{center}
\begin{tabular}{ c|c c c c c c}
\hline \hline
 & Linexp & Besselexp & Gaussian & Lognormal & Exponential & Top-Hat \\
\hline
Type 1 bias: & 216 &  179 &  108 &  145 &  323 &  294 \\
Type 2 bias:  & 5  &   2  &   8  &   1  &  10  &  12 \\
\hline
\end{tabular}
\end{center}
\end{table}

For sub-optimal $\chi^2_{SFH}$ bias, we find the distance $d = (\chi^2_{SFH}|_{min(\chi^2_{SFH})} - \chi^2_{SFH}|_{min(\chi^2_{SED})} )$ for each galaxy with each SFH family. This distance denotes the difference between the best $\chi^2_{SFH}$ possible in the basis and the $\chi^2_{SFH}$ corresponding to the best-fit SED in the basis. If the latter quantity is much worse than the former, we say that a galaxy has a bias due to sub-optimal $\chi^2_{SFH}$. We find that this is best quantified by the condition $d > 0.4 dex$. We show the number of these biases for each family in Table \ref{table:gofgor}, finding much lower rates of occurrence and that the Top-Hat family shows the highest occurrence of this behaviour.

\newpage

\end{document}